\documentclass[a4paper]{article}
\usepackage[utf8]{inputenc}
\usepackage{graphicx}
\usepackage{times}
\usepackage{natbib}
\usepackage{paralist}
\usepackage{booktabs}
\usepackage{subcaption}
\usepackage{multirow}
\usepackage{amsmath}
\usepackage{todonotes}

\setuptodonotes{
    backgroundcolor=blue!10, 
    bordercolor=none, 
}

\usepackage{hyperref}
\hypersetup{
    colorlinks=true,
    linkcolor=black,
    filecolor=blue,      
    urlcolor=blue,
    citecolor=red,
}
\usepackage{soul}
\usepackage{pgf}
\usepackage{tikz}
\usetikzlibrary{arrows,automata}
\usepackage{smartdiagram}
\usesmartdiagramlibrary{additions}
\usepackage{xspace}
\usepackage[acronym]{glossaries}
\usepackage{nicefrac}
\usepackage{authblk}

\newacronym{osm}{OSM}{Open Street Map}
\newacronym{dem}{DEM}{Digital Elevation Model}
\newacronym{dtm}{DTM}{Digital Terrain Model}
\newacronym{gtfs}{GTFS}{General Transit Feed Specification}
\newacronym{matsim}{MATSim}{Multi-Agent Transport Simulation}
\newacronym{ivabm}{IV-ABM}{Infrastructure Victoria Activity Based Model}
\newacronym{vitm}{VITM}{Victorian Integrated Transport Model}
\newacronym{tasha}{TASHA}{Travel Activity Scheduler for Household Agents}
\newacronym{vista}{VISTA}{Victorian Integrated Survey for Travel and Activity}
\newacronym{abs}{ABS}{Australian Bureau of Statistics}
\newacronym{csv}{CSV}{Comma Separated Values}
\newacronym{sa1}{SA1}{Statistical Area level 1}
\newacronym{sa2}{SA2}{Statistical Area level 2}
\newacronym{sa3}{SA3}{Statistical Area level 3}
\newacronym{sa4}{SA4}{Statistical Area level 4}
\newacronym{lga}{LGA}{Local Government Area}
\newacronym{atom}{AToM}{\textbf{A}ctivity-based and agent-based \textbf{T}ransport model \textbf{o}f \textbf{M}elbourne}
\newacronym{pt}{PT}{Public Transport}
\newacronym{atap}{ATAP}{Australian Transport Assessment and Planning}
\newacronym{thtv}{THTV}{Typical Hourly Traffic Volumes}
\newacronym{mnl}{MNL}{Multi-Nomial Logit model}
\newacronym{bn}{BN}{Bicycle Network}
\newacronym{mape}{MAPE}{Mean Absolute Percentage Error}
\newacronym{abm}{ABM}{Agent-Based Model}

\newcommand{\pkg}[1]{{\small{\texttt{\textbf{#1}}}}}

\title{\textbf{A}ctivity-based and agent-based \textbf{T}ransport model \textbf{o}f \textbf{M}elbourne (AToM): an open multi-modal transport simulation model for Greater Melbourne}

\author[a]{Afshin Jafari\thanks{Corresponding Author. Email address: afshin.jafari@rmit.edu.au. Postal address: RMIT University, GPO Box 2476, Melbourne VIC 3001.}}
\author[b,c]{Dhirendra Singh}
\author[a]{Alan Both}
\author[a]{Mahsa Abdollahyar}
\author[a]{Lucy Gunn}
\author[a]{Steve Pemberton}
\author[a]{Billie Giles-Corti}

\affil[a]{School of Global, Urban and Social Studies, RMIT University}
\affil[b]{School of Computing Technologies, RMIT University}
\affil[c]{Data61, CSIRO}

\date{\today}

\begin{document}

\maketitle

\begin{abstract}
    
    Agent-based and activity-based models for simulating transportation systems have attracted significant attention in recent years.
    Few studies, however, include a detailed representation of active modes of transportation---such as walking and cycling---at a city-wide level, where dominating motorised modes are often of primary concern.
    This paper presents an open workflow for creating a multi-modal agent-based and activity-based transport simulation model, focusing on Greater Melbourne, and including the process of mode choice calibration for the four main travel modes of driving, public transport, cycling and walking.
    The synthetic population generated and used as an input for the simulation model represented Melbourne's population based on Census 2016, with daily activities and trips based on the Victoria's 2016-18 travel survey data.
    The road network used in the simulation model includes all public roads accessible via the included travel modes.
    We compared the output of the simulation model with observations from the real world in terms of mode share, road volume, travel time, and travel distance.
    Through these comparisons, we showed that our model is suitable for studying mode choice and road usage behaviour of travellers.
    
\end{abstract}

\section{Introduction}\label{sec:intro}

    Computer-based transport simulations have been used for more than five decades to inform transport system management and decision-making~\citep{mcnally_four-step_2007}.
    The traditional approach to building transport simulation modelling was to divide the system's behaviour into four main steps: 
    \begin{inparaenum}[(i)]
        \item \emph{trip generation} (how many trips?);
        \item \emph{trip distribution} (between which zones?);
        \item \emph{modal split} (using which travel modes?); and
        \item \emph{assignment} (via which routes?).
    \end{inparaenum}
    Although four-step models have paved the way for the widespread use of simulation in planning for transport systems, a key limitation is their inability to associate trips to individuals and consequently to capture the heterogeneous behaviours of travellers, interactions between them, and inter-dependencies between different components of the transport system (e.g., infrastructure, congestion, travellers' mode and route preferences and trip chains) \citep{rasouli_activity-based_2014}.
    
    Activity-based modelling of transport systems addresses many of the four-step models' shortcomings through a dis-aggregated approach involving modelling individuals and their trips, activities, and heterogeneous decision-making and behaviours \citep{mcnally_activity-based_2007, rasouli_activity-based_2014}.
    Activity-based models take a bottom-up approach and simulate the individual behaviour of each entity of the system, the interactions between entities as well as with the environment ~\citep{kagho_agent-based_2020}. 
    From this perspective, the disaggregated approach of activity-based modelling is congruent with \glspl{abm} that 
     are computational models of heterogeneous agents and their interactions within their environment that can be used for experimenting with different possible scenarios \citep{gilbert_agent-based_2021}.
     Thus, significant benefit could be derived by joining the two approaches to capture both heterogeneous travel plans and complex interactions between travellers \citep{tajaddini_recent_2020,horl2020open}.
    
    \gls{matsim} is an open-source transport simulation toolkit that provides this link between agent-based and activity-based models and has become popular for large-scale transport models over the last decade \citep{horni_multi-agent_2016,hagerAgentbasedModelingTraffic2015}.
    \gls{matsim} is designed and optimised for large-scale simulations, which makes it a suitable option for city-wide models. 
    Notable examples of the development of large-scale \gls{matsim} models include  Switzerland \citep{bosch2016ivt}, Singapore \citep{erath2012large}, Melbourne \citep{infrastructure_victoria_model_2017}, while more recent models include those for Paris \citep{horl2020reproducible} and Berlin \citep{ziemke_matsim_2019}. 
    Furthermore, \gls{matsim} has been used to model different aspects of the transport system including \gls{pt} \citep{rieser_modeling_2016}, cycling \citep{ziemke2019bicycle}, and novel concepts such as shared mobility \citep{becker_assessing_2020} and shared autonomous electric vehicles \citep{muller_vienna_2021}.
    
    Another important trend in modelling the transport system of cities is the move towards multi-modal agent-based and activity-based models rather than the traditional approach of modelling only car and \gls{pt}.
    For example, \citet{oh_assessing_2020} analysed the impact of automated mobility-on-demand services using a multi-modal agent-based model for Singapore.
    The agent-based model \citet{chapuis_multi-modal_2018} model for flood emergency management for Hanoi, Vietnam, is another example of large-scale multi-modal simulation models.
    Despite the rise in multi-modal simulation models of cities, developing such models is a complicated and involved process and omitted to date is a flexible process for creating large-scale active transport simulation models using open data and tools.
    
    In this paper, we present our work on building a \textit{large-scale simulation model of the transport system for Greater Melbourne}, Australia.
    Our model is based on \gls{matsim} simulation toolkit and is the first multi-modal  \textit{calibrated and open}\footnote{We note that the first calibrated activity-based MATSim model for Melbourne was MABM~\citep{infrastructure_victoria_model_2017}. Ours is the first calibrated multi-modal model that is also \textit{open}. We also use more recent census and VISTA data than MABM.} activity-based MATSim model for Melbourne. 
    With this model, we aim to fill the gap of multi-modal open large-scale models in the literature and to use it as a baseline model for future simulation studies of Melbourne's transport system with a focus on active transport (i.e., walking, cycling and \gls{pt}).
    Furthermore, the complete workflow of producing the model as well as the tools we developed as part of the process are available on GitHub\footnote{\url{https://github.com/matsim-melbourne}} with the aim of addressing the need for flexible tools and processes for creating large-scale simulation models for active transport.
    
    The remainder of this paper is laid out as follows. Section~\ref{sec:background} provides an overview of the key concepts in building activity-based transport models using the \gls{matsim} simulation toolkit.
    Section~\ref{sec:method} describes our workflow and the key tools and methods used to develop the simulation model for Melbourne.
    The calibration process of the model and evaluation of the calibrated scenario are discussed in Section~\ref{sec:outputAnalysis}.
    Finally, in Section~\ref{sec:discussion} we discuss how this model could be used to help inform decision-making for the transport system in Melbourne, and the applicability of the framework to other cases and potential future steps of the model.

\section{Background}\label{sec:background}

    The three main building blocks of an \gls{abm} are: 
    \begin{inparaenum}[(i)]
        \item a synthetic population of heterogeneous agents;
        \item their environment; and 
        \item a way for agents to interact with each another and their environment \citep{wall2016agent}.
    \end{inparaenum}
    For transport system \glspl{abm}, the synthetic population is a list of travellers, their attributes, and their travel diaries.
    Different approaches to build synthetic populations for transport modelling are briefly reviewed in Section~\ref{sec:demand_background}.
    The environment for transport \glspl{abm} is typically the road network that the agents use to travel to their daily destinations. 
    Section~\ref{sec:supply_background} reviews recent studies for creating road network models for \glspl{abm}.
    Lastly, we used \gls{matsim} as our \gls{abm} simulation framework to model the transport system.
    Section~\ref{sec:matsim_background} briefly discusses \gls{matsim} and how it models the interaction amongst agents and their environment.

    \subsection{Synthetic population construction}\label{sec:demand_background}

    Synthetic population generation based on the activity-based modelling framework typically involves steps for generating a list of agents with their demographics, assigning activity patterns (i.e., activity chain or itinerary), and assigning locations to activities \citep{wang_improved_2021}.
    Over the years, a number of different methods have been developed to produce the synthetic population for activity-based and agent-based transport models.
    
    A widely used approach is to create a synthetic population based on probability distributions from travel surveys.
    For example, \gls{tasha}, a well-known travel demand generator calibrated for Greater Toronto Area \citep{roorda_validation_2008}, uses a joint probability distribution function for creating activity-based travel demands.
    In \gls{tasha}, population and demographics replicate Greater Toronto's transport survey, representing 4.5\% of the population.
    The survey was also used to create joint probability functions for different activity types, demographics, household structure, and trip schedules, resulting in 262 distributions that were used to generate activities.
    A similar probabilistic approach was also used to select the activity start time and duration for each activity.
    These functions were then used to generate the list of activities of each individual.
    Home and work locations in \gls{tasha} were given to the model as inputs and locations for other activities were assigned using entropy models based on distance, employment density, population density, and measures for other land-use types such as shopping mall floor space for the shopping activity \citep{roorda_validation_2008}. 
    
    More recently, machine learning techniques have been used to enhance synthetic population generation accuracy and flexibility \citep{koushik2020machine}.
    For example, \citet{hesam2021framework} used techniques from  machine learning along with econometric techniques and proposed a hybrid framework for creating activities and travel diaries using a cohort-based synthetic pseudo panel engine to model.
    Similarly, a k-means clustering algorithm was used by \citet{allahviranloo2017modeling} to cluster activities based on trip attributes and to synthesize activity chains.

    \citet{both2021activity} proposed an algorithm for creating the synthetic population for the Greater Melbourne area using a combination of machine learning, probabilistic and gravity-based approaches.
    In their algorithm, each synthetic individual was assigned a demographic profile (e.g., age, gender) consistent with the census population at the \gls{sa2} geospatial boundary,\footnote{Demographic distributions were matched to \gls{abs} Census 2016 at the \gls{sa2} geospatial boundary, which conceptually represents a community of on average 10,000 persons.} a home location (i.e., a valid street address in that \gls{sa2}), and a daily travel plan comprising a sequence of activities (at particular locations and times of the day) connected by travel legs (using particular modes, e.g., driving, \gls{pt}) consistent with the travels observed for persons of that demographic profile in the \gls{vista} 2012-18 travel survey data \citep{department_of_transport_victorian_2018}.
            
    Five destination types of home, work, education, commercial, and park were included in the algorithm.
    The destinations of different types were distributed across Greater Melbourne based on the Vicmap Address database by the Victorian government\footnote{ \url{https://discover.data.vic.gov.au/dataset/address-vicmap-address}} containing 2,932,530 addresses and their Mesh Block (MB) land use categories.
    MB is the smallest geographical area defined by \gls{abs} and residential MBs have a dwelling count of approximately 30 to 60 in urban areas.\footnote{\url{https://www.abs.gov.au/ausstats/abs@.nsf/Lookup/by\%20Subject/1270.0.55.001~July\%202016~Main\%20Features~Mesh\%20Blocks\%20(MB)~10012}} 
    Both et al.'s algorithm also assigns a travel mode to each trip based on the starting region's probability to be used for the location assignment process \citep{both2021activity}.
    
\begin{figure}
    \centering
    \begin{tabular}{cc}
    	\includegraphics[width=0.45\columnwidth]{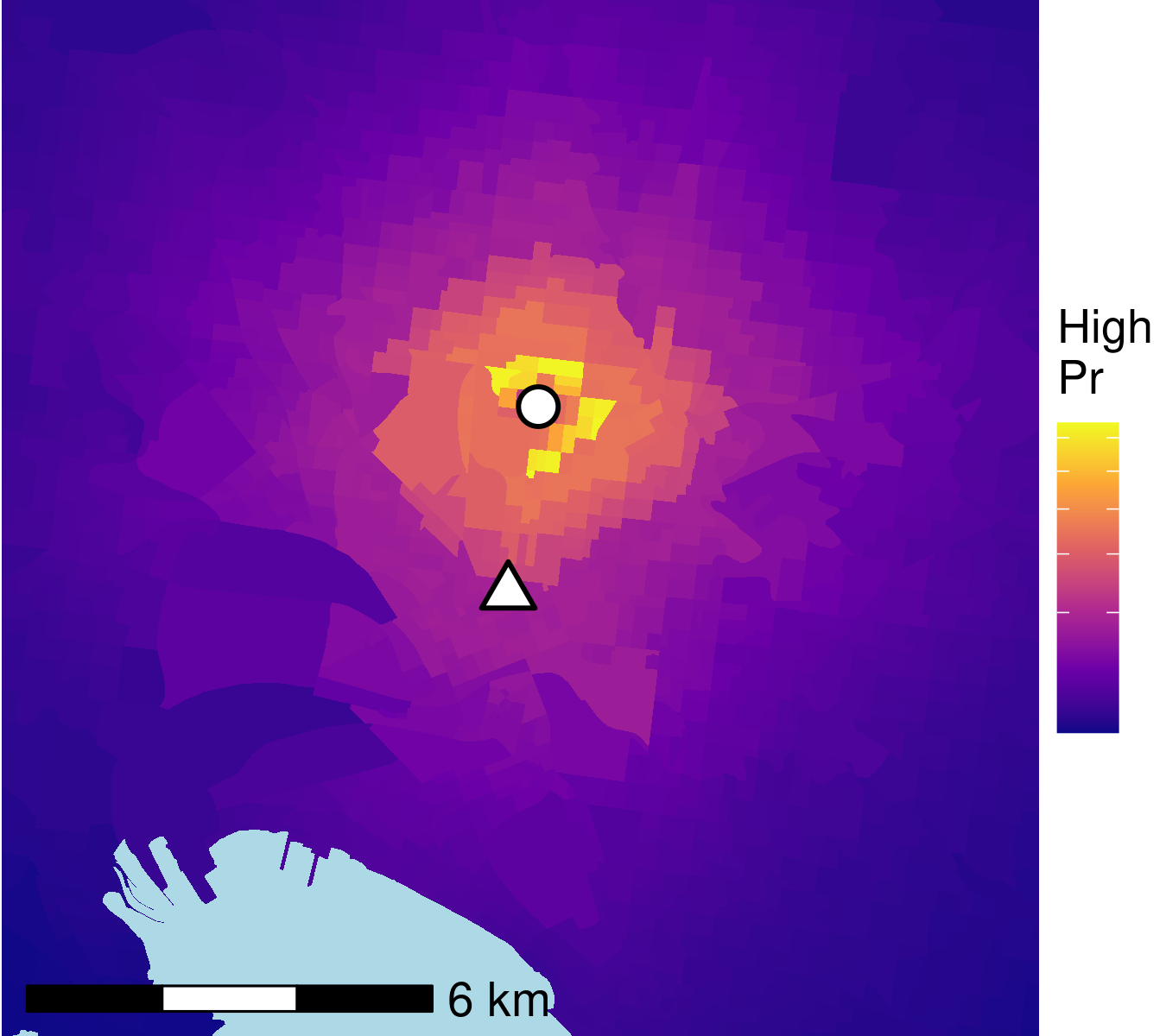} &
    	\includegraphics[width=0.45\columnwidth]{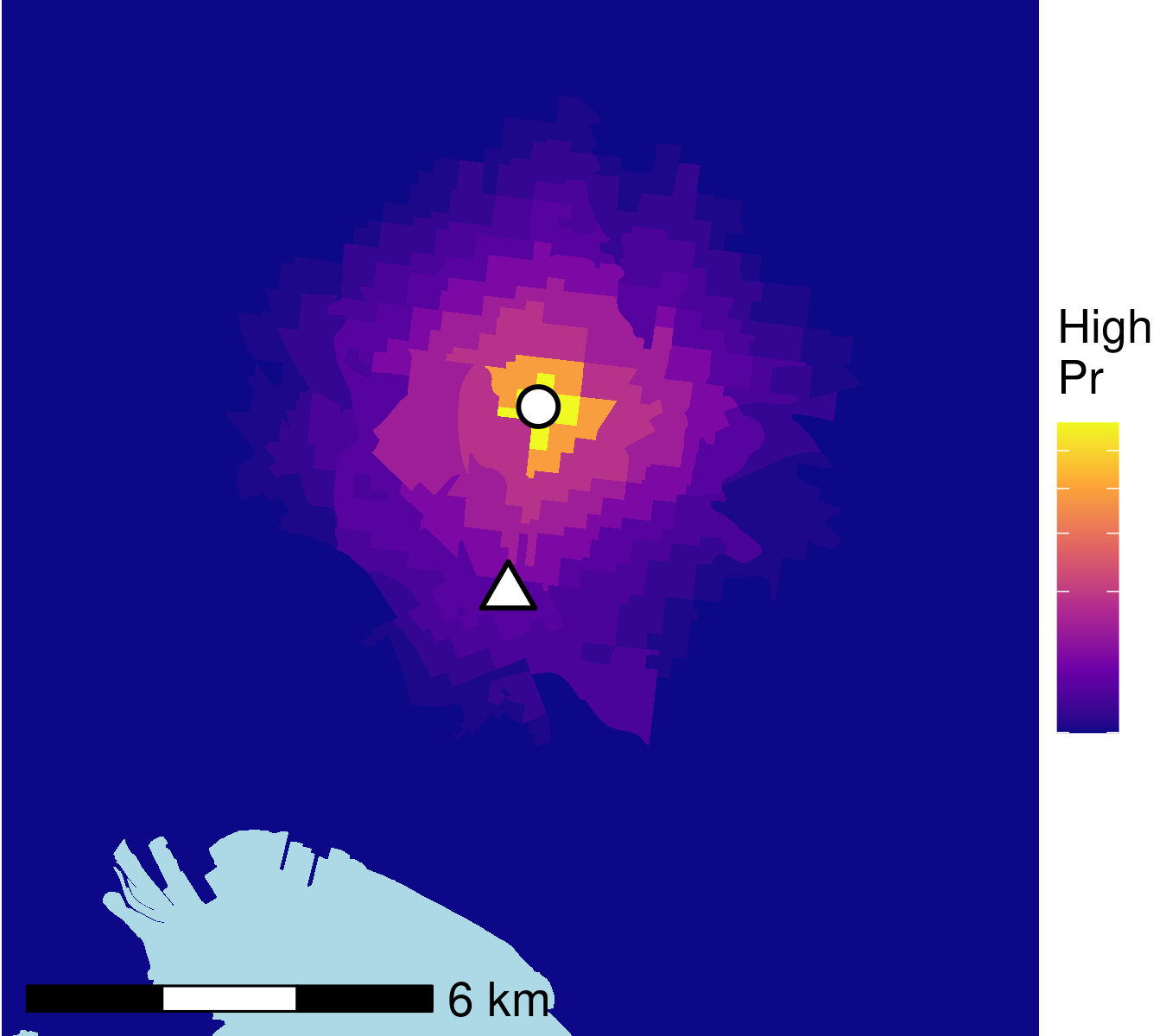}\\
    	(a) & (b)\\
    	\includegraphics[width=0.45\columnwidth]{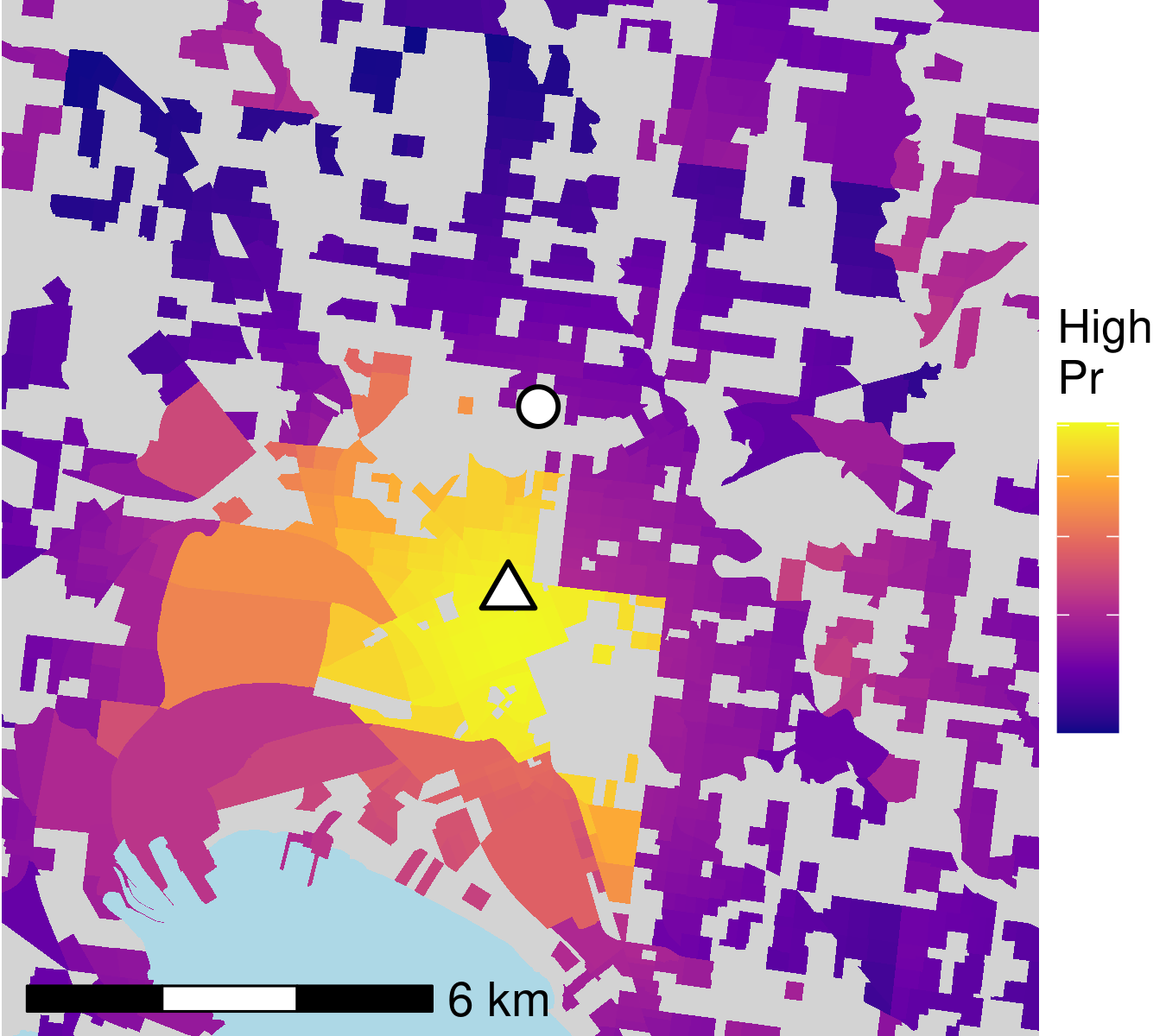} &
    	\includegraphics[width=0.45\columnwidth]{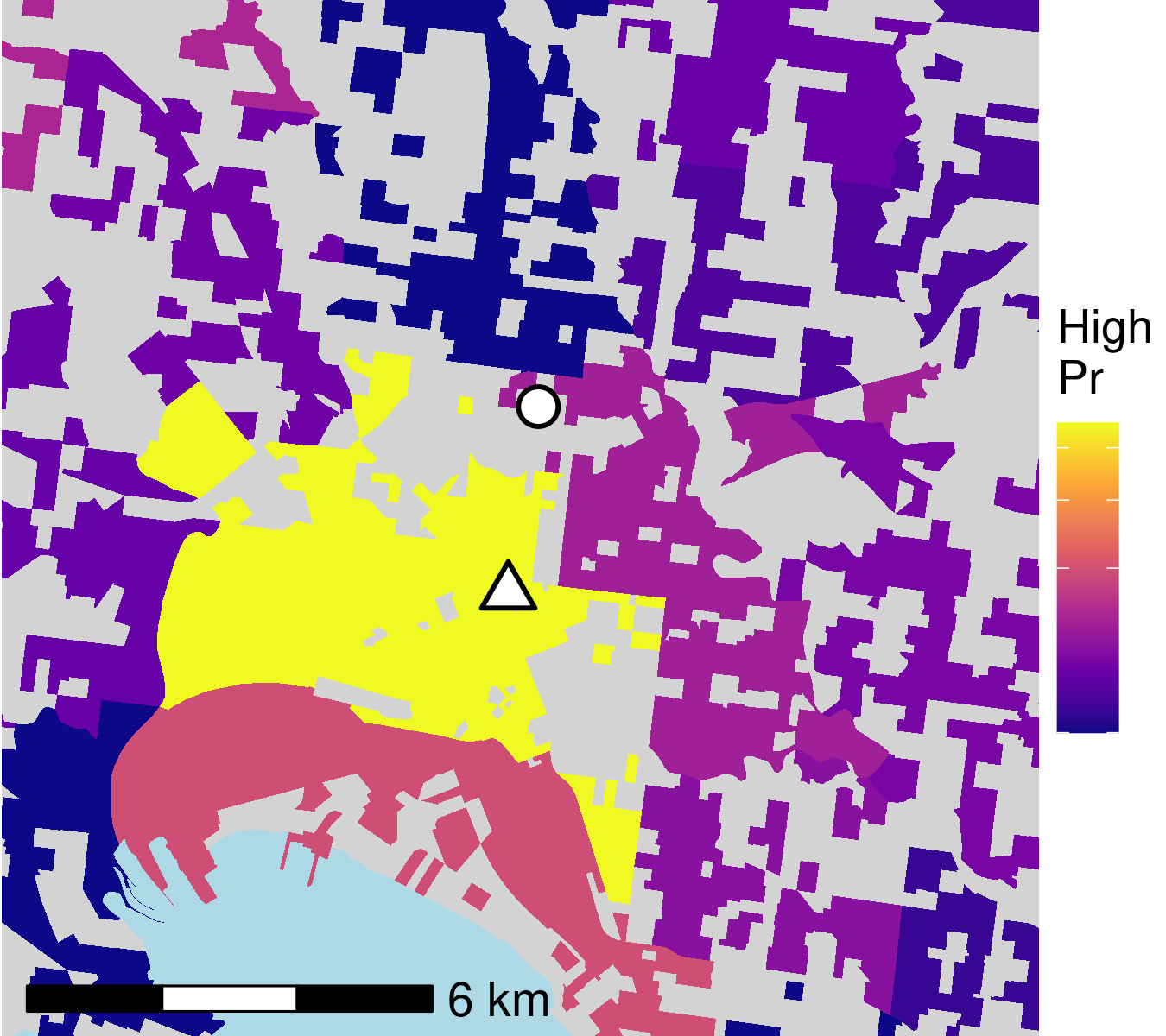}\\
    	(c) & (d)\\
    	\includegraphics[width=0.45\columnwidth]{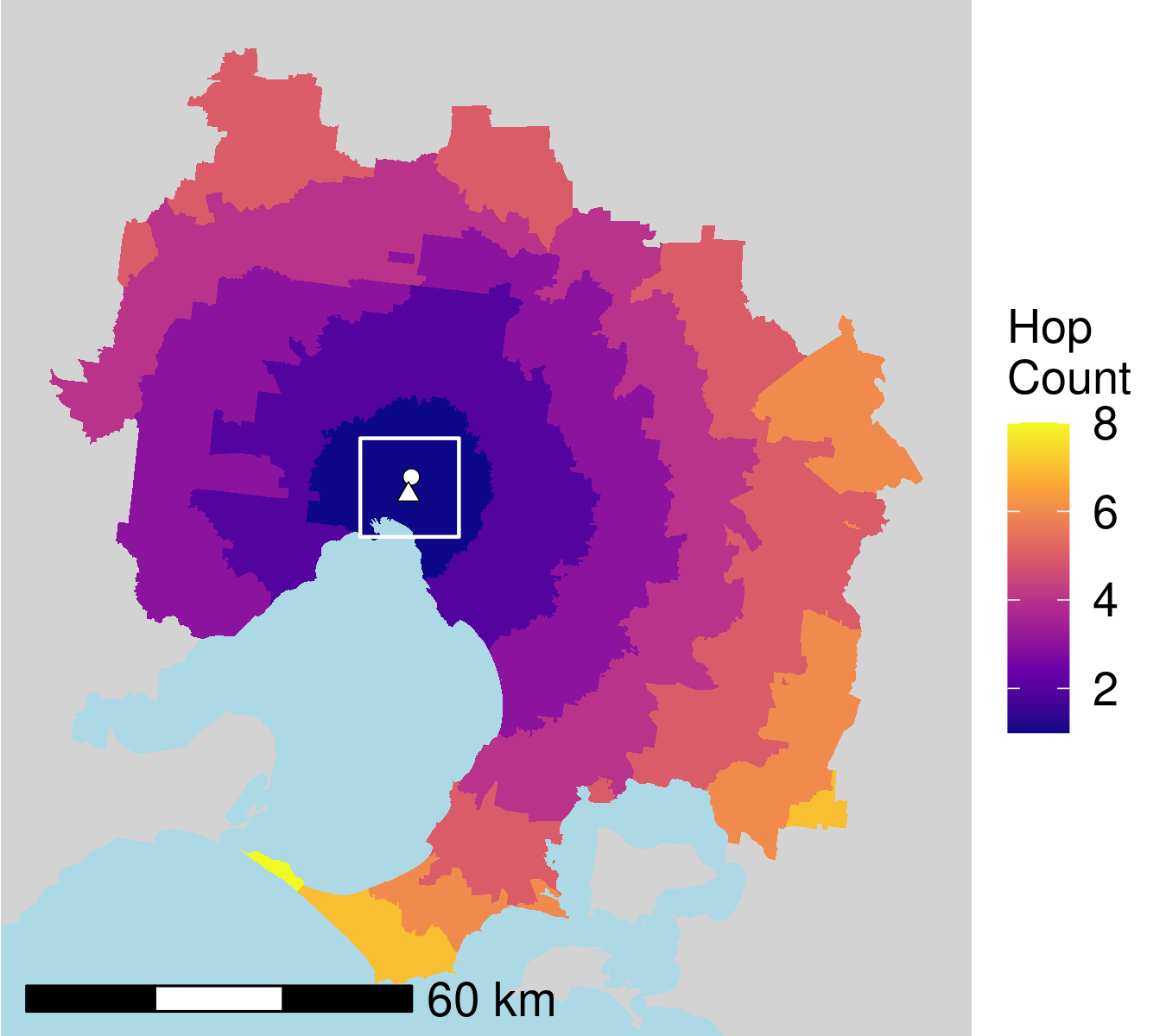} &
    	\includegraphics[width=0.45\columnwidth]{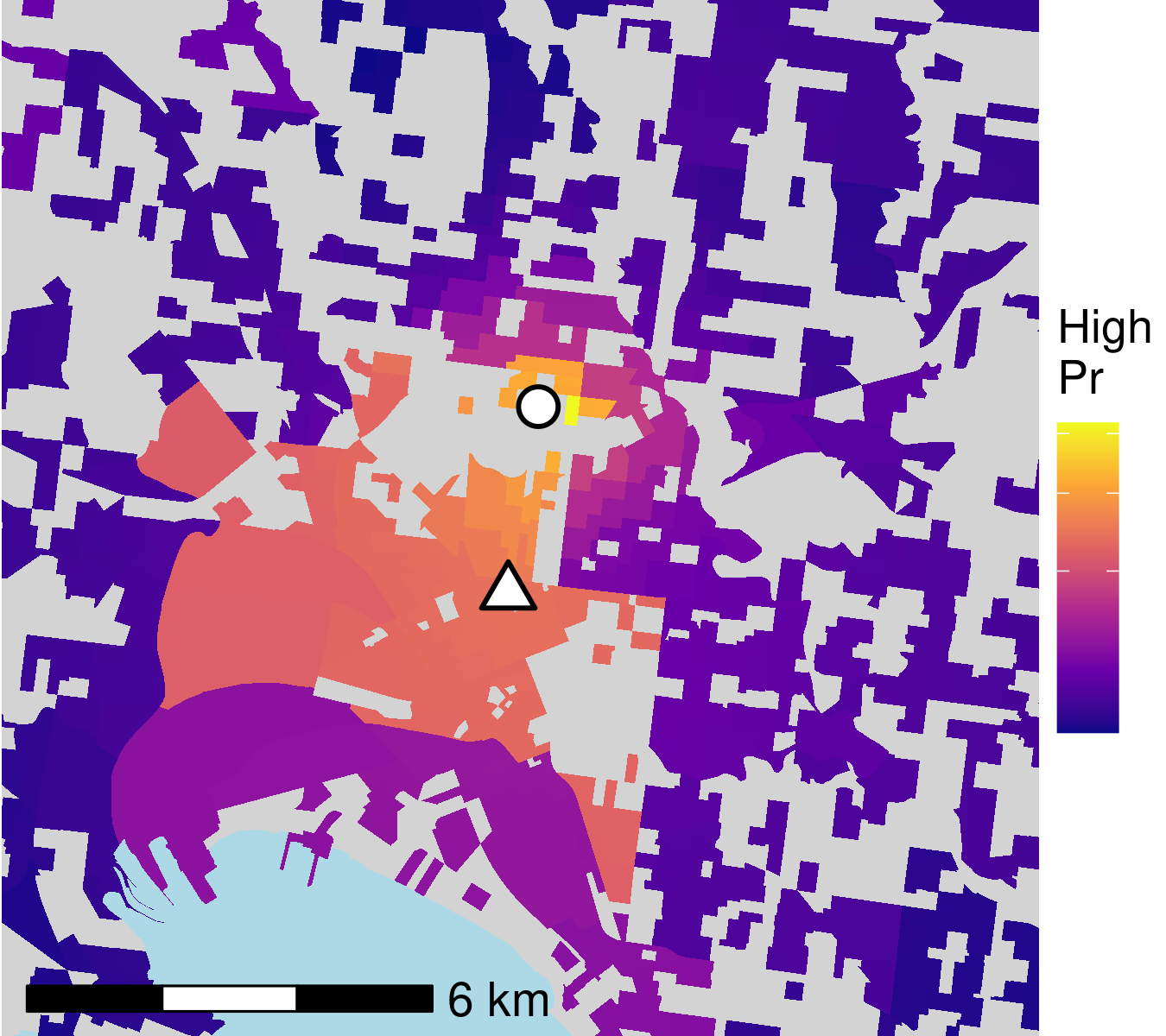}\\
    	(e) & (f)\\
    \end{tabular}
    \caption{Selecting next region for a cycling trip from home (circle) to work (triangle) showing: region selection probability (Pr) for local and global distance distributions (a and b), region selection probability (Pr) for local and global destination attraction (c and d), number of trips (hop count) that would be reasonably required to reach home (e), and combined region likelihood (f) (source: \citep{both2021activity}).}
    \label{fig_selectingNextRegion}
    \end{figure}

    When assigning locations to activities, the assigned transport mode, the distance traveled to the destination, and the destination itself for the activity were required to account for local variation while also conforming to global distributions.
    To ensure this was the case, for each SA1 region, values were calculated based on the \gls{vista} travel survey.
    New locations were chosen sequentially for each agent, with the restriction that agents start and finish at home.
    Transport mode was chosen first, so that the candidate regions can be filtered down to the ones likely for that mode, based on the number of trips remaining to get home.
    This was to ensure that the final trip home will not be unreasonably long. 
    The remaining regions were then ranked based on their distance and how likely the agent would be to choose the region based on the local distance distribution and the attractiveness of that region for the specified location type. 
    Additionally, global distance distribution and destination attraction were considered to ensure that the synthetic population's overall trip length and destination choice reflected that of the \gls{vista} travel survey.
    Figure \ref{fig_selectingNextRegion} illustrates how the distance distribution and transport mode probabilities, and destination type probabilities were combined to create the probabilities needed to select the next region.
    
    \subsection{Road network model construction methods}\label{sec:supply_background}
    
        One of the main inputs of the transport simulation models is the road network description.
        This not only indicates the location of the road infrastructure that agents can travel through, but also assesses its quality and usage specifying, for example, road capacity and speed limits, and what modes are allowed on particular roads.
        In other words, the synthetic population creates the transport system demand, while the road network provides the supply.
        
        In recent years, \gls{osm} has become a useful and reliable source of transport infrastructure information for use in modelling transport systems.
        \gls{matsim} has built-in functionality to convert raw \gls{osm} extracts into  \gls{matsim} readable transport networks for car traffic \citep{zilske_openstreetmap_2015}.
        There have been efforts to expand \gls{matsim}'s \gls{osm} converter over the last few years and as a result a number of complementary tools have been developed.
        For example, \citet{poletti_public_2016} developed a tool called \textit{pt2matsim} to find and add \gls{pt} routes to the  \gls{matsim} network based on \gls{gtfs} -- a common format used globally for \gls{pt} schedules and associated geographic information.
        \citet{ziemke2019bicycle} further extended  \gls{matsim}'s network converter tool to incorporate bicycle-relevant attributes, including slope, surface type, and bicycle-specific infrastructure, creating a detailed network for bicycle traffic simulation.
        
        To create a road network for a city including all major transport modes (driving, \gls{pt}, walking, and cycling), one approach is to combine the \gls{matsim} tools listed above.
         \citet{jafari_building_2022} recently proposed an open and standalone algorithm that integrates these steps and automates the process of building a city-wide network.
        The network generation process starts by extracting road geometries from \gls{osm} and converting them to a set of links and nodes.
        Their algorithm includes components for adding road elevation from a \gls{dem}, adding a \gls{pt} network from \gls{gtfs}, simplifying the network to make it suitable for large-scale simulation experiments, and finally creating a \gls{matsim} readable network to be used for simulation.
        \citet{jafari_building_2022} showed that their network simplification algorithm creates a significantly reduced network compared with the network generated using the algorithm proposed by \citet{ziemke2019bicycle}, with minimal loss of detail needed for active transport modelling, yet significant gains in simulation run-time performance.

    \subsection{MATSim framework for activity-based transport simulations}\label{sec:matsim_background}
        
\begin{figure}[h]
            \centering
            \begin{tikzpicture}[
    ->, >=latex, shorten >=1pt, shorten <=1pt, auto,
    node distance=2.5cm, line width=1.5pt, draw=black!50!white
  ]
  \tikzstyle{every state}=[
    rectangle, rounded corners=0.2cm, text=black, 
    minimum height=1.2cm,
    text width=1.6cm, align=center
  ]

  \node[state]         (A)              {Input Demand};
  \node[state]         (B) [right of=A] {Execution};
  \node[state]         (C) [right of=B] {Scoring};
  \node[state]         (D) [below = of $(B)!0.5!(C)$, yshift=1.2cm] {Replanning};
  \node[state]         (E) [right of=C] {Analysis};

  \path (A) edge []  node {} (B)
        (B) edge []  node {} (C)
        (C) edge []  node {} (E)
        (C) edge [bend left]  node {} (D)
        (D) edge [bend left]  node {} (B);
        
\end{tikzpicture}
            \caption{The MATSim process loop}
            \label{fig:matsimLoop}
        \end{figure}
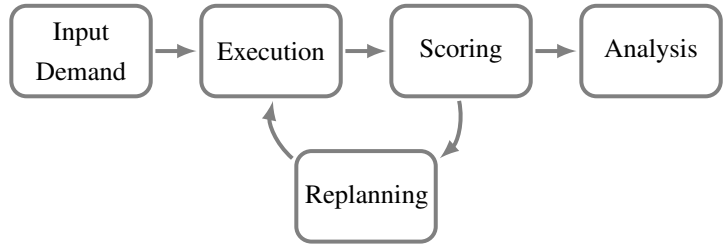

    	\gls{matsim} follows a co-evolutionary optimisation algorithm to determine how the supply from the road network is to be used by the demand from the  synthetic population \citep{horni_multi-agent_2016}.
    	Figure~\ref{fig:matsimLoop} illustrates \gls{matsim}'s optimisation process known as the \gls{matsim} loop.
    	The process starts with each agent obtaining an travel and activity plan for the day, i.e., the initial \emph{input demand} coming from the synthetic population.
    	Then agents perform their plans simultaneously, (i.e., \textit{execution}) and travel to their destinations based on \gls{matsim}'s queue-based traffic mobility simulator using the road network.
    	All executed plans get scored based on a utility function, (i.e., \textit{scoring} (Equation~\ref{eq:totalUtil})).
        Next, each agent remembers the score of a limited number of previous iterations' plans, and based on these, selects its travel plan for the next iteration, i.e., \textit{re-planning}.
        
    	During the re-planning process, a given percentage of agents modify their chosen plan following different strategies such as randomly varying departure times, travel mode, and routes.
    	The iterative process of execution, scoring, and re-planning gets repeated until the rate of increase of the average score of all selected and simulated plans across the  synthetic population plateaus, i.e., tends to zero. 
    	The output of the simulation from the final iteration is then used for further examination (i.e., \textit{analysis}) \citep{horni_multi-agent_2016}. 
    
        The score of an executed plan, which represents how well an agent's simulated day goes compared to its desired plan, is calculated in \gls{matsim} as follows.
        
        \begin{equation}\label{eq:totalUtil}
            S_{plan}=\sum_{p=1}^{N}S_{act,p}+\sum_{q=1}^{N}S_{trav,mode(q)},
        \end{equation}
        
        where $S_{plan}$ is the total score of the plan, $S_{act,p}$ is the positive score of an activity $p$, $S_{trav,mode(q)}$ is the negative score of travel on trip leg $q$ and, and $N$ is the total number of activity destinations in the agent's plan. 
        Trip $q$ is the trip that follows activity $p$, and assuming two activities are connected by a single travel leg, we have $N$ activity destinations and $N$ trips to travel to them. 
        A minimal equation to calculate the score of activity $p$, $S_{act,p}$, is shown in Equation~\ref{eq:activityUtil}.
        
        \begin{equation}\label{eq:activityUtil}
            S_{act,p}=S_{dur,p}+S_{late.ar,p},
        \end{equation}
        
        where $S_{dur,p}$ is the score of performing the activity $p$ for the duration of $dur$ and $S_{late.ar,p}$ is the score of late arrival to activity $p$. 
        
        In a simple case, the score of travel for leg $q$, the second component in  Equation~\ref{eq:totalUtil}, is equal to:
        
        \begin{equation} \label{eq:disUtil}
            S_{trav,q} = asc_{mode(q)} + \beta_{trav,mode(q)}.t_{trav,q} + \beta_{m}.\Delta m_{q} + (\beta_{d,mode(q)} + \beta_{m}.\gamma_{d,mode(d)}).d_{trav,q} . 
        \end{equation}
        			
        In Equation~\ref{eq:disUtil}, $asc_{mode(q)}$ represents a mode-specific constant, $\beta_{trav,mode(q)}$ is the direct \textit{marginal utility of time} spent travelling by mode $q$ and $t_{trav,q}$ is the travel time to activity $p$ with mode $q$. 
        $\beta_{m}$ is the \textit{marginal utility of money}, $\Delta m_{q}$ is the change in the monetary budget caused by fares or toll, $\beta_{d,mode(q)}$ is the \textit{marginal utility of distance} and $\gamma_{d,mode(d)}$ is the monetary cost per kilometre for each travel mode, i.e., \textit{monetary distance rate}. 
        The distances between activities is denoted by $d_{trav,q}$. 

\section{\acrshort{atom}: model development workflow and calibration}\label{sec:method}

\begin{figure}[h]
        \centering
        \includegraphics[width=\textwidth]{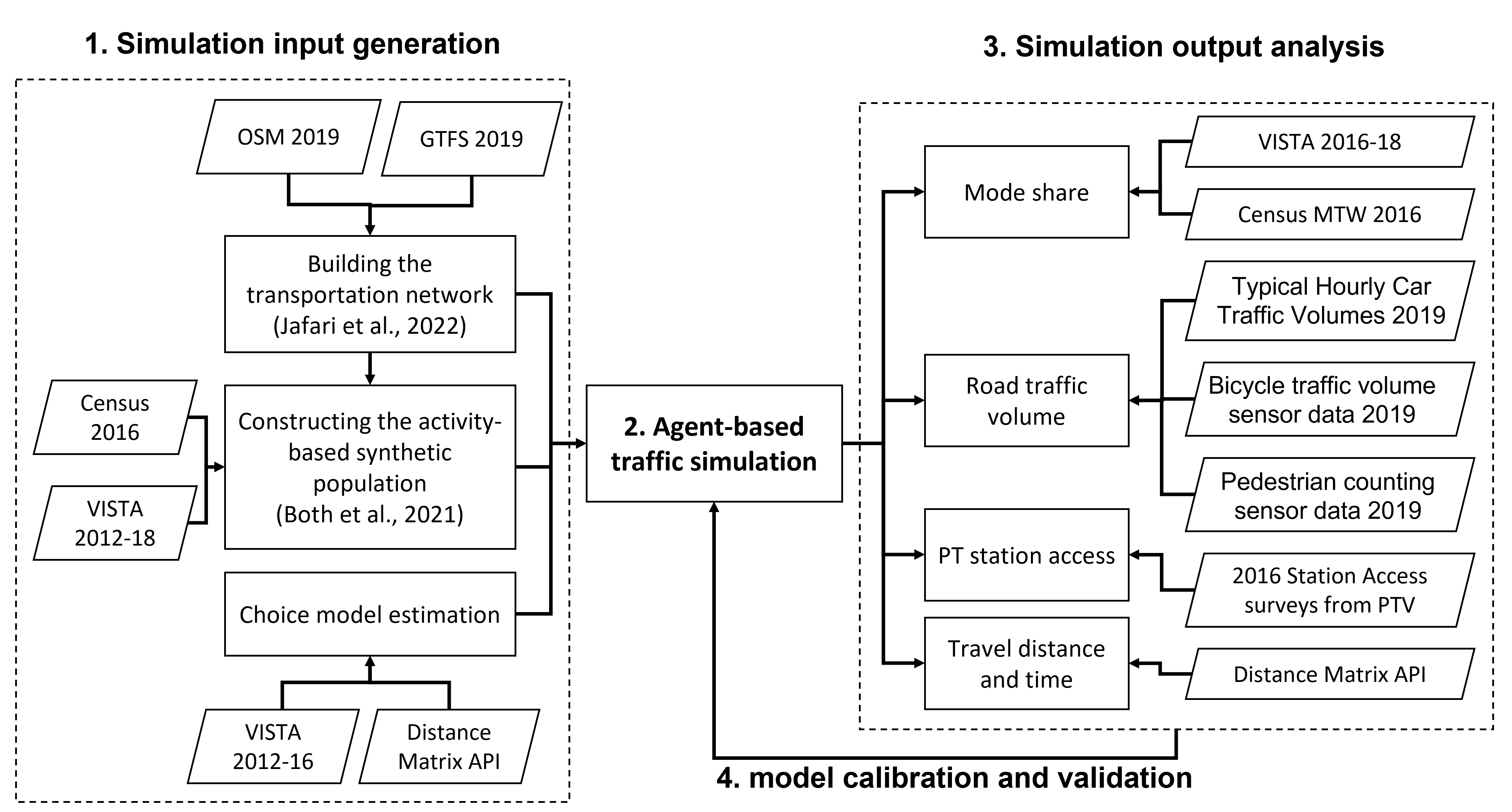}
        \caption{The model development workflow overview}
        \label{fig:pipeline}
    \end{figure}

    Figure~\ref{fig:pipeline} provides an overview of the \gls{atom} development workflow.
    The process started with building the road network (Section~\ref{method:network}).
    Synthetic population generation was the next step of the workflow (Section~\ref{method:demand}). 
    Although the algorithm used to create the synthetic population did not rely on the road network as an input, we used the network nodes as its optional input and snapped the activity destinations to their nearest network node.
    This was to ensure that the activity locations generated by the algorithm were joined to and were directly accessible via the road network. 
    Estimating the model parameters was the next component of the workflow that is described in Section~\ref{method:choice_model}.
    These three components were then used as the simulation inputs for the agent-based traffic simulation model as detailed in Section~\ref{method:model}.
    The simulation output analysis was the next step where simulated mode share, road traffic volume, and travel distance and time were compared to real-world observations. 
    The process of running the simulation model, analysis and comparison of the simulation outputs, and adjustment of the model to better fit the observed data, i.e., the calibration loop, is covered in Section~\ref{sec:outputAnalysis}.
    
    As Figure~\ref{fig:pipeline} shows, different waves of the \gls{vista} data set were used by different components in our workflow.
    The most recent \gls{vista} wave, i.e., 2016--18, has the most recent sample of travellers in Melbourne.
    However, it has the limitation that the destination locations were reported at \gls{lga} level, with some \glspl{lga} such as the City of Wyndham and the City of Melton having a land area of more than 500$km^2$.
    Whereas, in earlier versions of \gls{vista}, for years 2012 to 2016, destination locations were reported at \gls{sa1} according to Australian Statistical Geography Standard (ASGS), an area with an average population of 400 people.\footnote{\url{https://www.abs.gov.au/ausstats/abs@.nsf/Lookup/by\%20Subject/1270.0.55.001~July\%202016~Main\%20Features~Statistical\%20Area\%20Level\%201\%20(SA1)~10013}}
    In this paper, for the model parameter estimation where higher destination location accuracy was desired, we used \gls{vista} 2012-16, whereas for mode choice calibration where the most recent data was desired, we used the 2016-18 wave.
            
    \subsection{Simulation input generation}\label{method:inputGen}
        As explained in Section~\ref{sec:matsim_background}, a minimum \gls{matsim} model requires a synthetic population for the traveller agents, a road network model of the study area, and a set of parameters (e.g., marginal utility of time and money) forming MATSim's evolutionary optimisation scoring function.
        In this section, we describe our process for creating these three inputs.
    
        \subsubsection{Building the transportation network (Network generator tool)}\label{method:network}
            
\begin{figure}[h]
                \centering
                \begin{subfigure}[b]{0.45\textwidth}
                	\includegraphics[width=\textwidth]{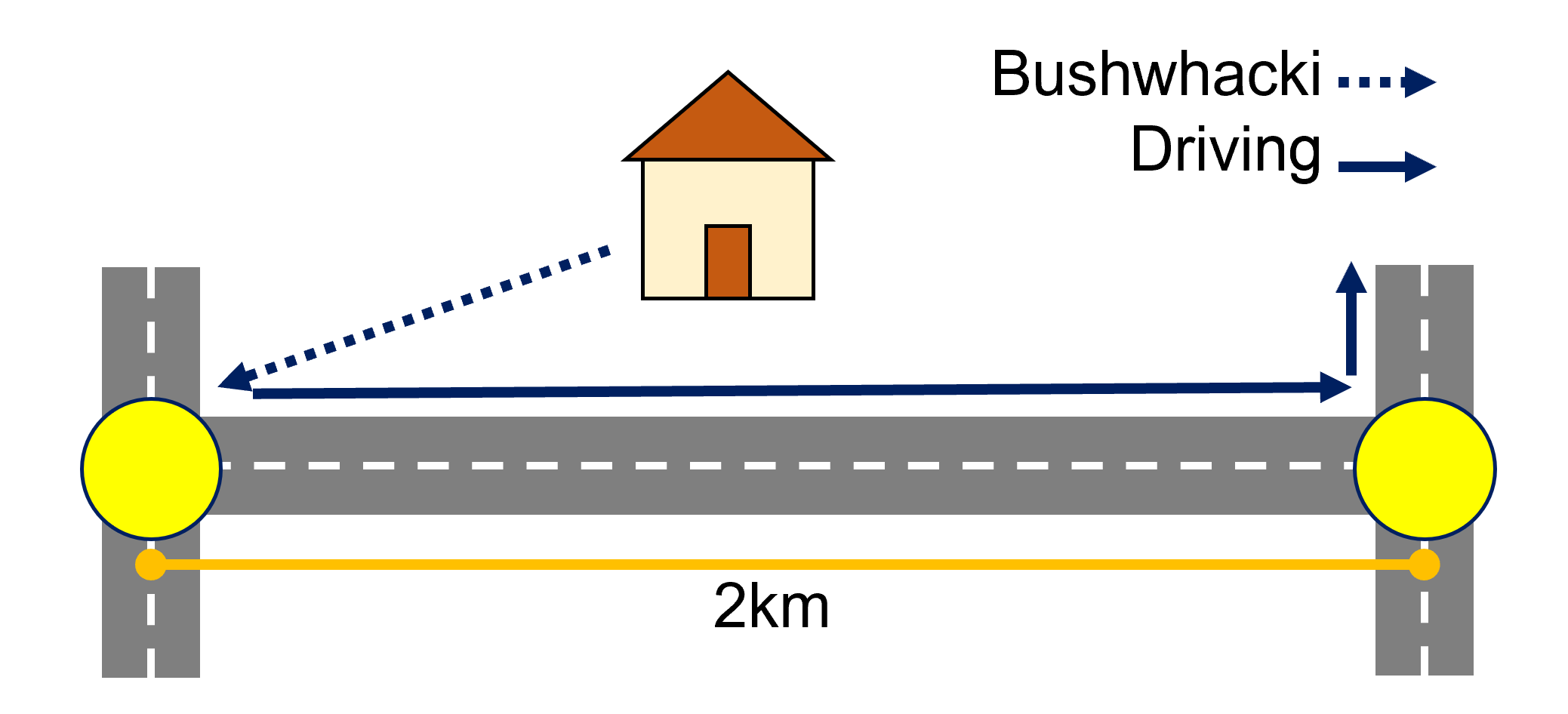} 
                	\caption{Without 500m access points}
                	\label{fig:bushwhackingBefore}
                \end{subfigure}
                \begin{subfigure}[b]{0.45\textwidth}
                	\includegraphics[width=\textwidth]{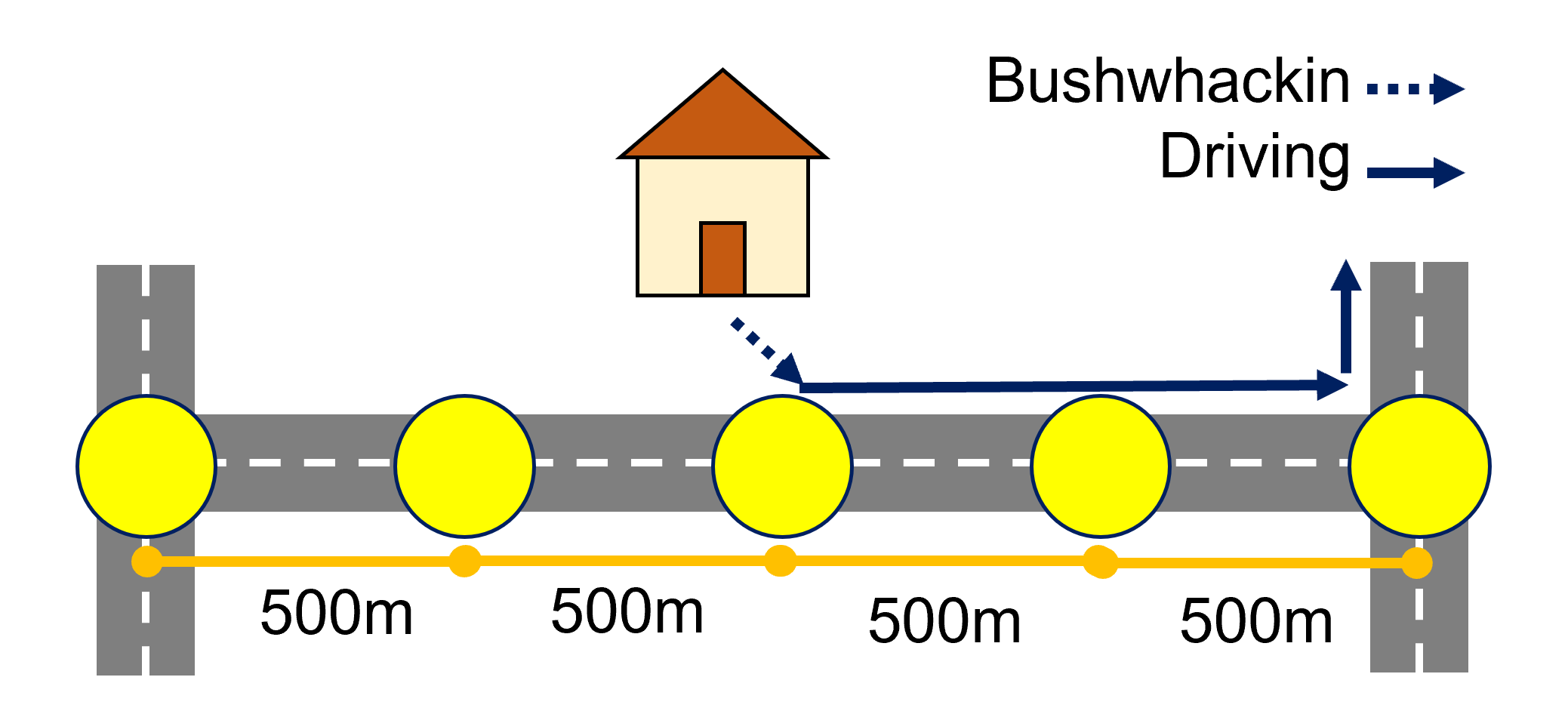} 
                	\caption{With 500m access points}
                	\label{fig:bushwhackingAfter}
                \end{subfigure}
                \caption{A schematic illustration of a car traveller entering traffic from the link's start node in MATSim with and without 500m access points}
                \label{fig:bushwhacking}
            \end{figure}

            The \gls{osm} extract for Greater Melbourne for October 2019 was used to create a \gls{matsim} compatible network for the Greater Melbourne using the algorithm proposed in \citet{jafari_building_2022}.
            The resulting network is in the form of a set of links representing road segments and nodes at every road break point, i.e., intersection, roundabout, or road access point.
            In \gls{matsim}, vehicles can only enter the traffic from the start node of a link.
            This could cause a considerable amount of travelling on non-existing roads for long links in \gls{matsim}, where agents must walk a considerable distance to get to a start node so that they can start travelling on the network using their designated mode (Figure~\ref{fig:bushwhackingBefore}).
            Therefore, to minimize this error, we divide any large road links (greater than a threshold length of 500m) in areas conducive to active modes (with a speed limit less than 60km/h, including a footpath, and permitting both walking and cycling), into several links no greater than the threshold length (Figure~\ref{fig:bushwhackingAfter}).
            In Melbourne, this filtration results in selecting local and residential roads where travellers can enter traffic from their driveways or parking lots, leaving out motorways and major roads where traffic can only enter at designated junctions. 
            
\begin{figure}[h]
                \centering
                \begin{subfigure}[b]{0.45\textwidth}
                	\includegraphics[width=\textwidth]{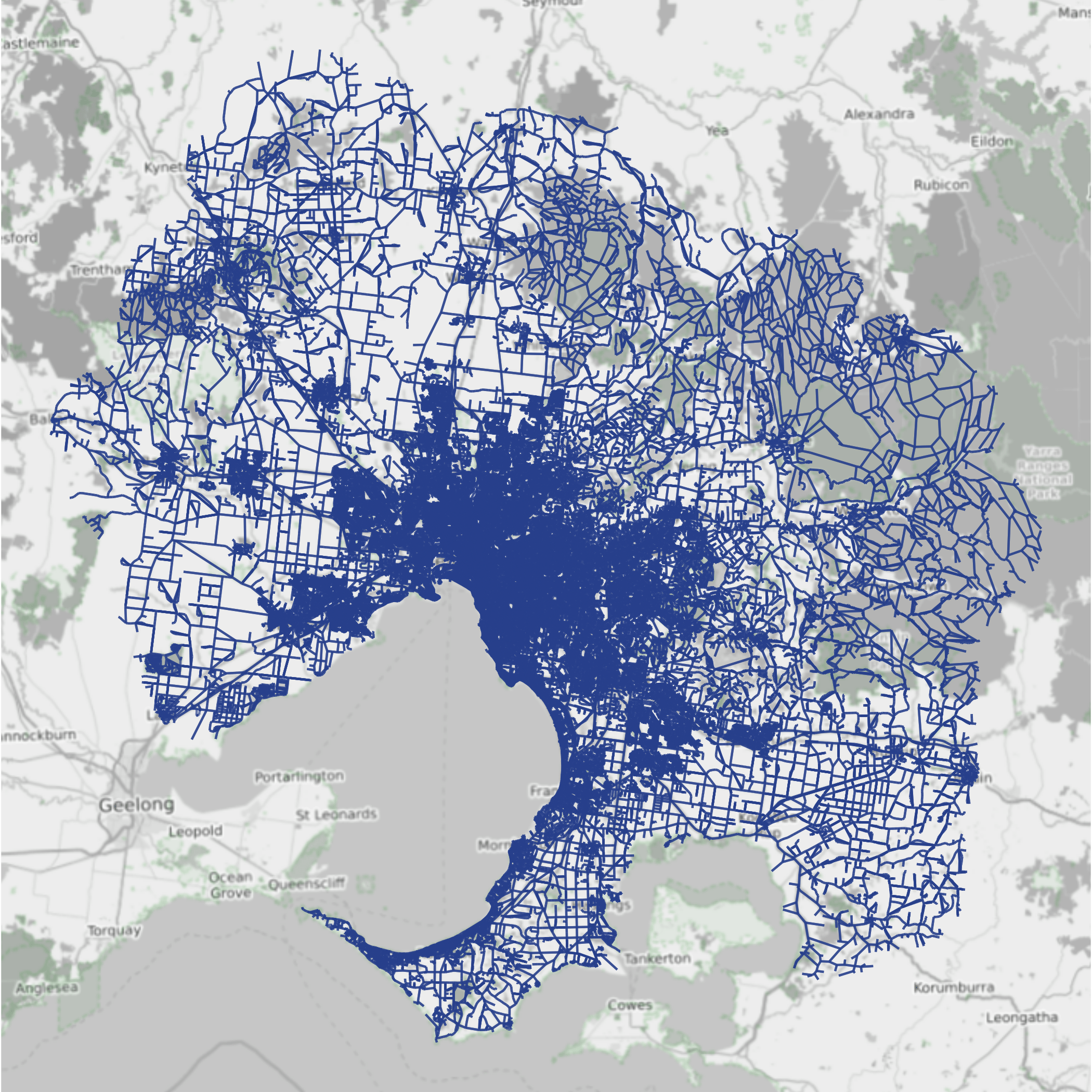} 
                	\caption{Road network (\gls{pt} excluded)}
                	\label{fig:roadNetwork}
                \end{subfigure}
                \begin{subfigure}[b]{0.45\textwidth}
                    \includegraphics[width=\textwidth]{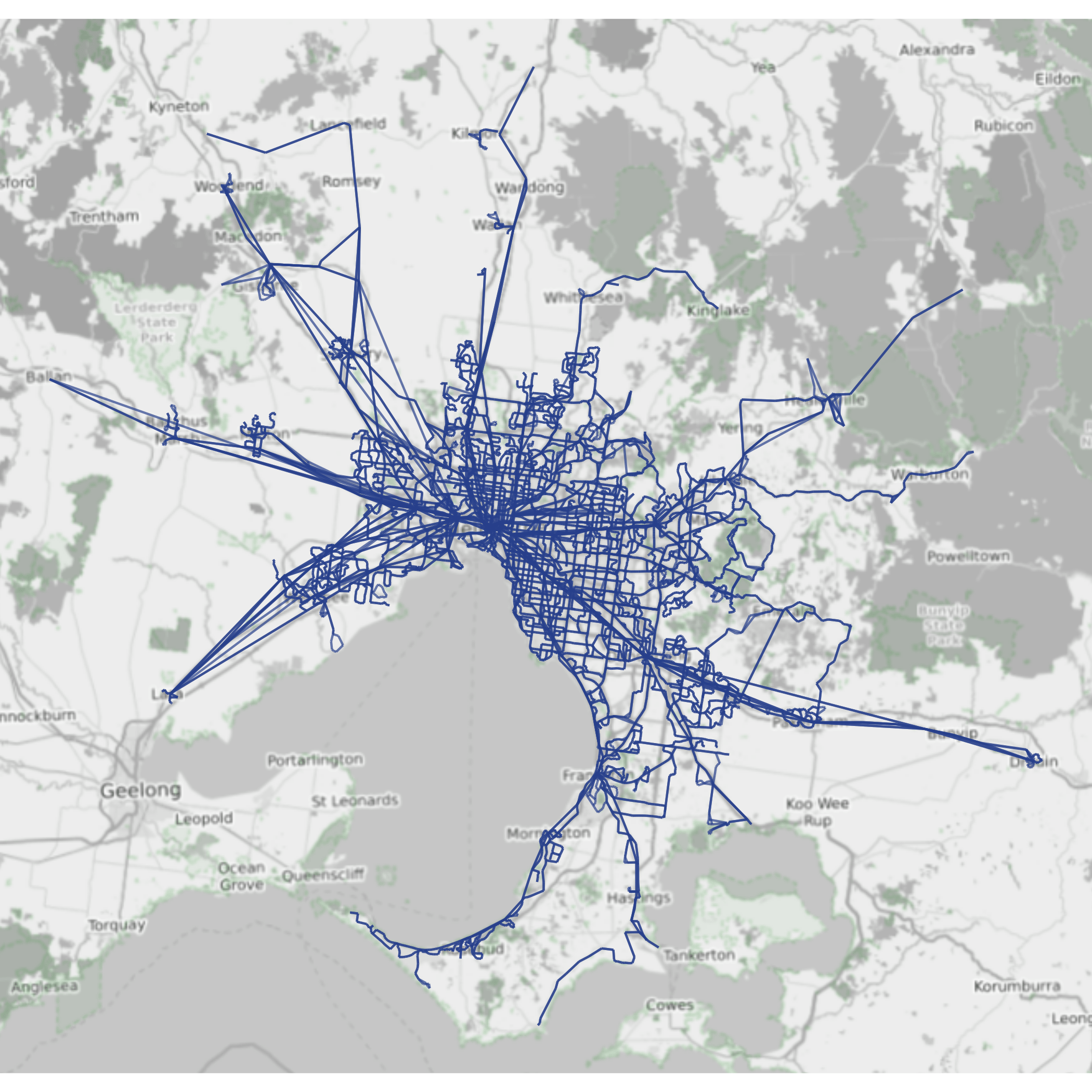} 
                	\caption{\gls{pt} network}
                	\label{fig:ptNetwork}
                \end{subfigure}
                \caption{Generated road network and \gls{pt} network for the study area}
                \label{fig:mmNetwork}
            \end{figure}

            Furthermore, a minimum link length of 20m was assumed to simplify the network for run time efficiency.
            This means connected links (i.e., road segments) with a length less than 20m were merged into a single node, resulting in a simpler representation of complex intersections and roundabouts.
            \citet{jafari_building_2022} argue that this simplification results in a significant decrease in simulation run-time without compromising the accuracy of model.
            Figure~\ref{fig:roadNetwork} depicts the generated road network for the study area.
            
            To simulate \gls{pt} trips, \gls{matsim} requires two additional inputs: one indicating the service lines, their stop locations, routes and schedules, and another giving a list of \gls{pt} vehicles with their types and carrying capacities.
            \gls{pt} fleet and service schedules and routes were created based on the
            \gls{gtfs} feed data for a time frame starting at 2019-10-11 and ending at 2019-10-17, downloaded from the OpenMobilityData website.\footnote{\url{https://transitfeeds.com/p/ptv}}
            \gls{pt} stop coordinates were snapped to the closest road network nodes to ensure all stops are accessible.
            The resulting \gls{pt} network is illustrated in Figure~\ref{fig:ptNetwork}.
        
        \subsubsection{Constructing the activity-based synthetic population (synthetic population generator tool)}\label{method:demand}
           
\begin{figure}[h]
                \centering
                \begin{subfigure}[b]{0.32\textwidth}
                	\includegraphics[width=\textwidth]{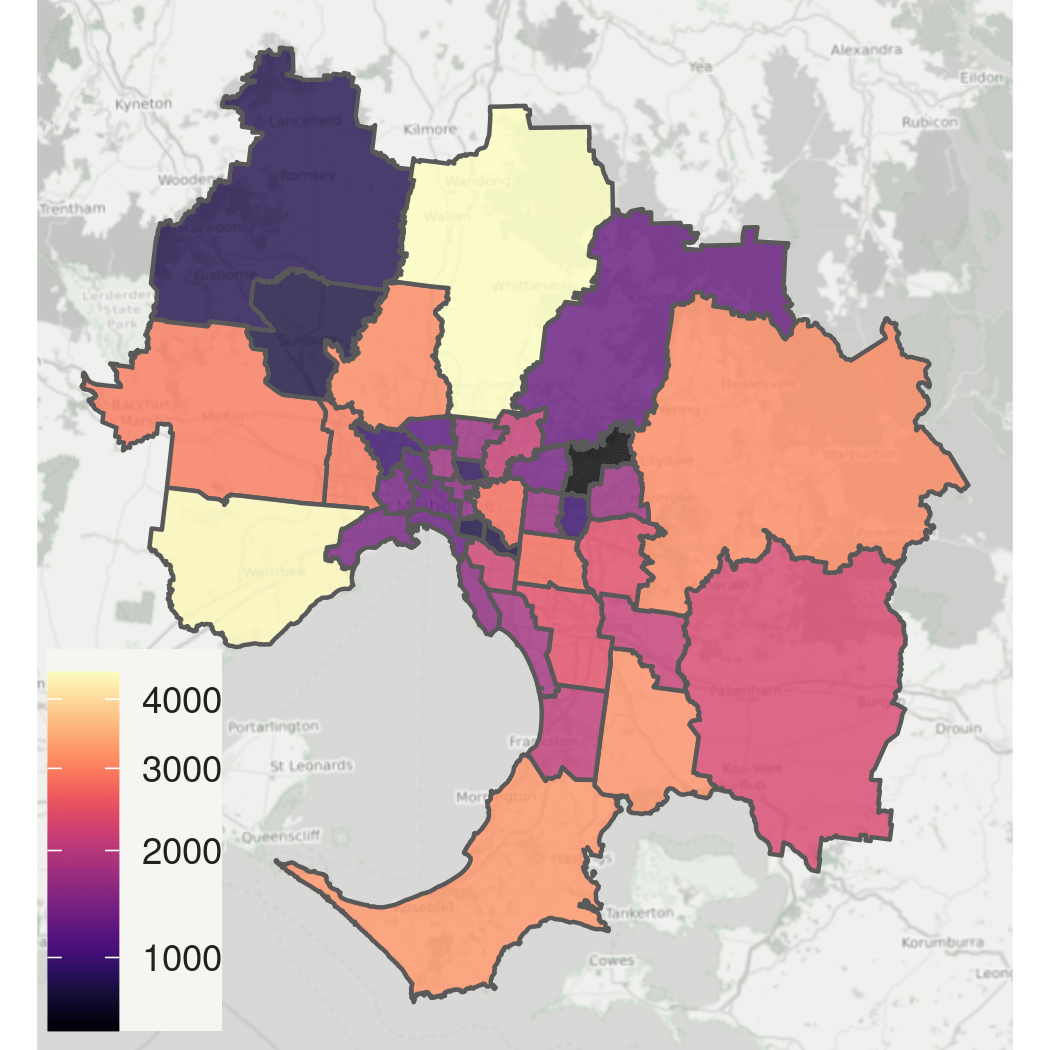} 
                	\caption{Home locations}
                	\label{fig:homeLocations}
                \end{subfigure}
                \begin{subfigure}[b]{0.32\textwidth}
                	\includegraphics[width=\textwidth]{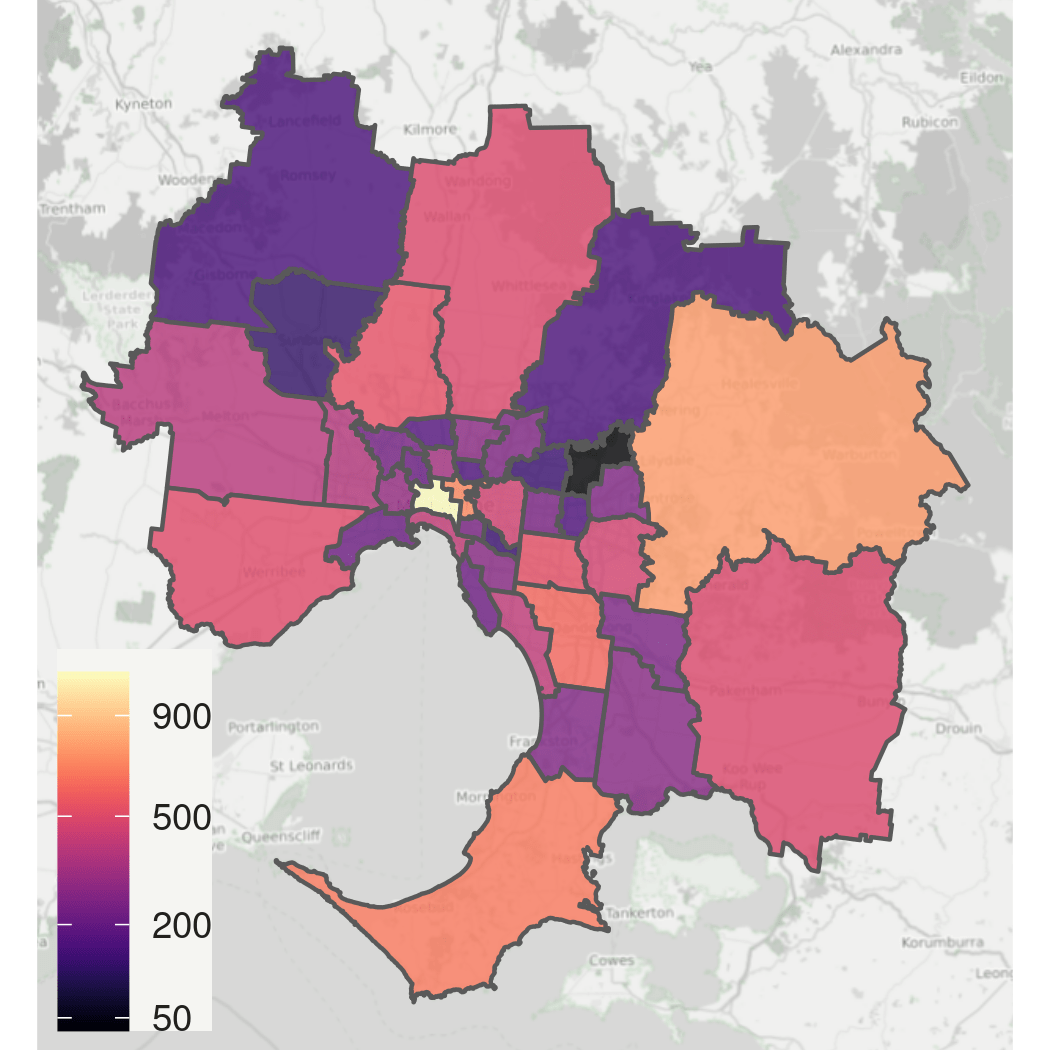} 
                	\caption{Work locations}
                	\label{fig:workLocations}
                \end{subfigure}
                 \begin{subfigure}[b]{0.32\textwidth}
                	\includegraphics[width=\textwidth]{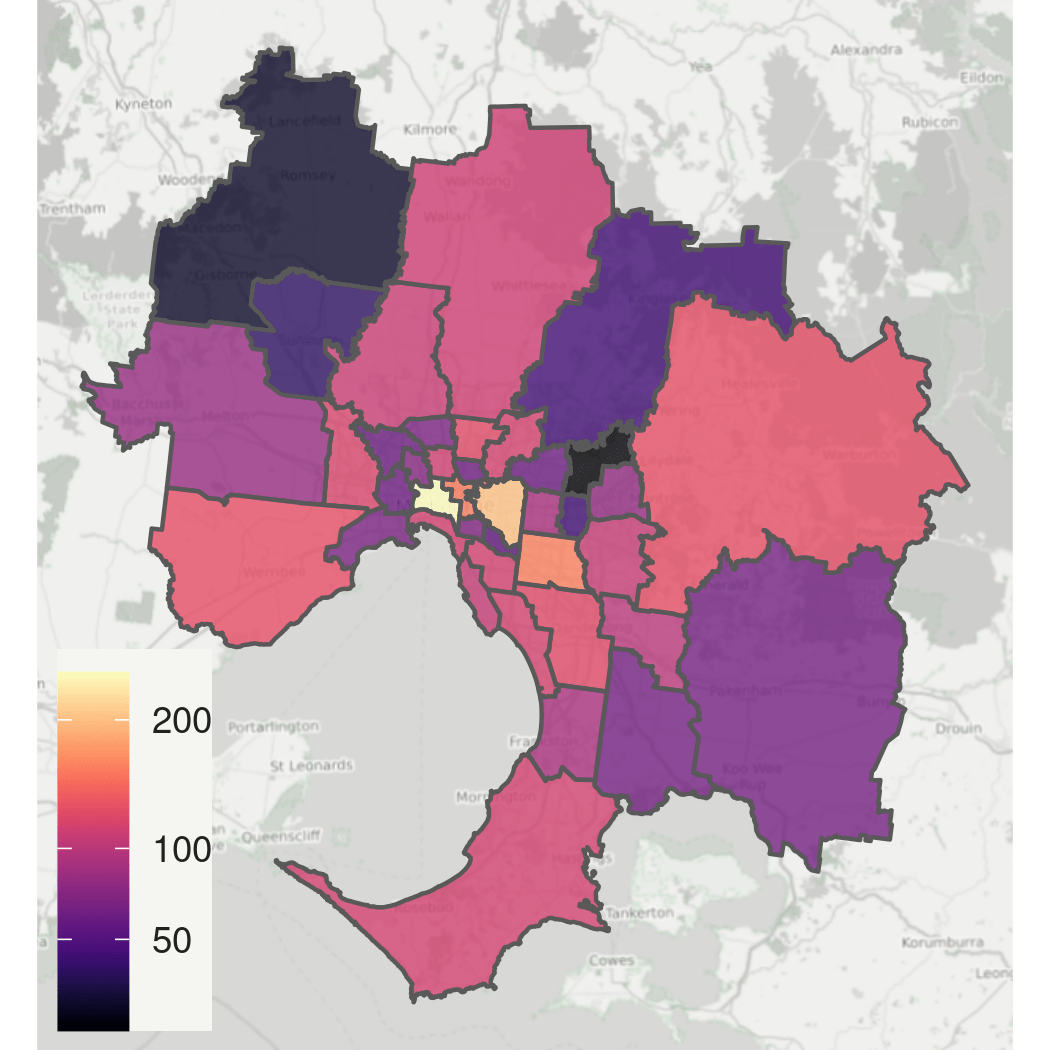} 
                	\caption{Education locations}
                	\label{fig:eduLocations}
                \end{subfigure}
                 \begin{subfigure}[b]{0.32\textwidth}
                	\includegraphics[width=\textwidth]{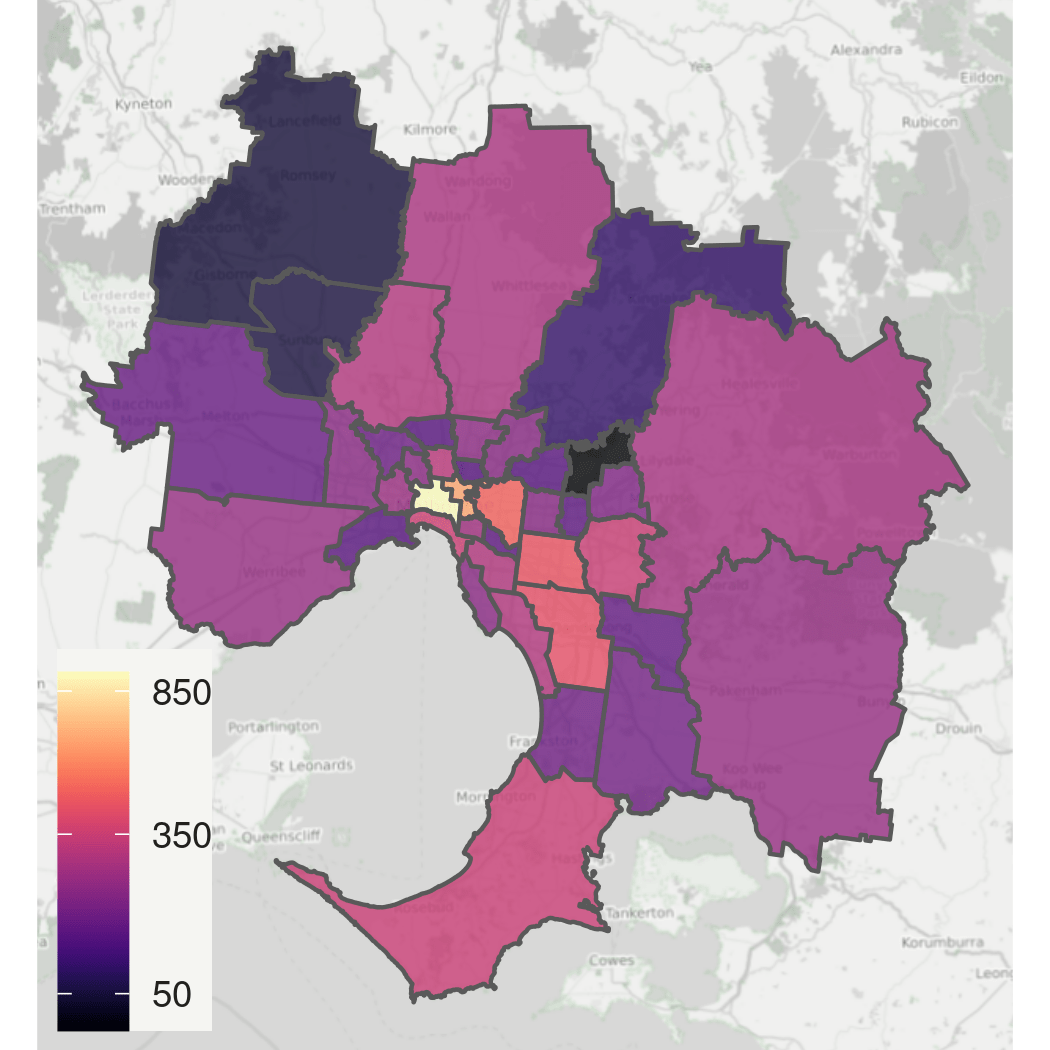} 
                	\caption{Commercial locations}
                	\label{fig:comLocations}
                \end{subfigure}
                 \begin{subfigure}[b]{0.32\textwidth}
                	\includegraphics[width=\textwidth]{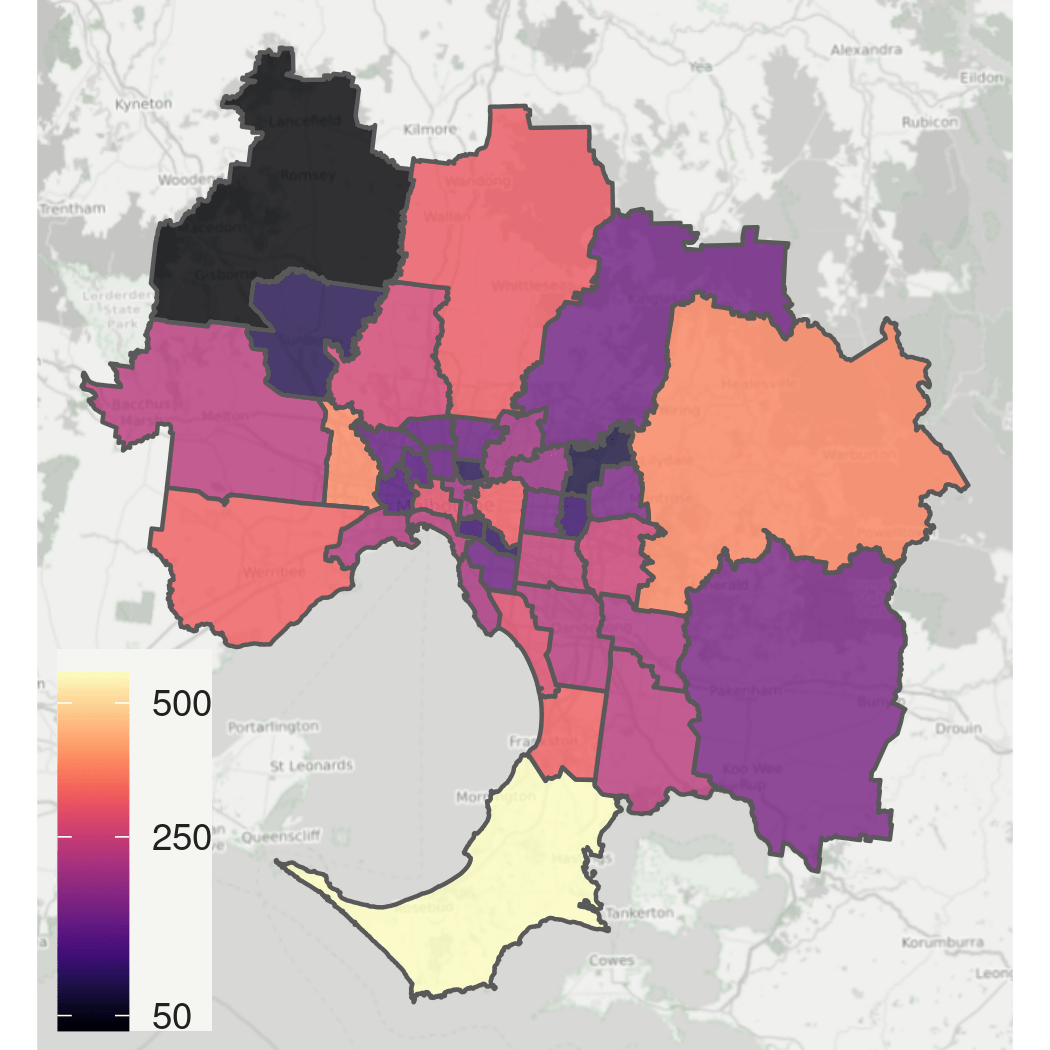}
                	\caption{Park locations}
                	\label{fig:parkLocations}
                \end{subfigure}
                \caption{Destination type location distribution aggregated at \gls{sa3} level}
                \label{fig:activityLocations}
            \end{figure}

            Using \cite{both2021activity}’s algorithm a synthetic population of individuals representative of the 10\% of the Greater Melbourne region population was generated. 
            The algorithm was implemented in R\footnote{\url{https://github.com/matsim-melbourne/demand}} and provides a convenience script to write the synthetic population out as a \gls{matsim} population XML file, which we used as-is in this work.
            The population generation algorithm ensures the overall travel destination locations, activity chains, and timing as well as individuals' profiles in the synthetic population are representative of the real population at the aggregated level. 
            Destination type location distribution aggregated at \gls{sa3} level across Greater Melbourne is shown in Figure~\ref{fig:activityLocations}.
            Interested readers are encouraged to see \citet{both2021activity} for a more comprehensive analysis of activity chains and timing.
                
            The synthetic population was divided into two sub-population groups of \textit{workers} and \textit{non-workers} based on whether they had a trip to work or not.
            These sub-population groups were used during the simulation to implement different behaviour change or innovation strategies for each as as explained in Section~\ref{method:model}.
            
        \subsubsection{Choice model estimation}\label{method:choice_model}

            We used \gls{vista} 2012-16 data as the main input to estimate \gls{matsim}'s utility function parameters.
            \gls{vista} trip records starting and finishing within the Greater Melbourne area and via one of the four travel modes of driving, \gls{pt}, walking, and cycling were selected.
            From the resulting set, commute trips from home to work or education (as primary destinations) were selected for further analysis, giving a sample of 14,959 from 92,725 total trips.
            Selection of mandatory commute trips to primary destinations was intended to minimise samples affected by factors such as personal goals that are highly relevant for recreational or social trips \citep{ramezani_shopping_2021}.
            \gls{vista} 2012-16 reports the origins and destinations of trips aggregated at the \gls{sa1} level. 
            Latitude and longitude coordinates of the \gls{sa1} centroids were considered as the coordinates of each trip origin and destination.
            
            The selected sample was used to estimate the \gls{matsim} mode choice parameters for Melbourne as discussed in Section~\ref{sec:background}.
            The first step was to specify the utility function (Equation~\ref{eq:disUtil}) for each travel mode alternative based on the model assumptions.
            We assumed the effect of distance to be fully captured through the travel time and cost components of the utility function, and therefore the marginal utility of distance was not considered for any of the four mode alternatives.
            Moreover, no monetary cost was assumed for walking and cycling trips, therefore, their utility function could be written as Equations~\ref{eq:Walk_Simple} and \ref{eq:Bike_Simple}, respectively.
            
            \begin{subequations}
        		\label{eq:SimpleChoiceModel}
        		\begin{align}
        		S_{trav, Driving} &=                    \beta_{trav, Driving} \times t_{trav, Driving} + \beta_{m} \times \Delta m_{Driving}, \label{eq:Car_Simple}   \\
        		S_{trav, PT}      &= asc_{PT}         + \beta_{trav, PT}      \times t_{trav, PT}      + \beta_{m} \times \Delta m_{PT}     , \label{eq:PT_Simple}    \\	
        		S_{trav, Walking} &= asc_{Walking}    + \beta_{trav, Walking} \times t_{trav, Walking}                                      , \label{eq:Walk_Simple}  \\
        		S_{trav, Cycling} &= asc_{Cycling}    + \beta_{trav, Cycling} \times t_{trav, Cycling}                                      . \label{eq:Bike_Simple} 
        		\end{align}
        	\end{subequations}
    	    
            For \gls{pt}, a trips-based constant fare was used to represent the monetary cost argument $\Delta m_{PT}$ of Equation~\ref{eq:PT_Simple}.
            According to \gls{vista} 2012-16, those who used \gls{pt} to get to work or education reported on average two \gls{pt} trips for their survey day.
            The daily \gls{pt} pass fare for Melbourne\footnote{Public transport fares in Melbourne vary based on the zones the person travels within or between, and whether the traveller has a daily, monthly, or even yearly \gls{pt} travel pass or is paying for each trip individually.
            For simplicity, we assumed \gls{pt} travellers to use the standard (zone1+2) daily travel pass.} in 2016 was \$7.80, giving an approximated average cost of $7.8/2=\$3.90$ per trip. 

            Lastly, a distance-based fuel consumption cost function $\Delta m_{Car}=\gamma_{d,Car} \times d_{trav,Car}$ was assumed for driving, where $\gamma_{d,Car}$ is the fuel consumption cost per km for an average vehicle and $d_{trav,Car}$ represents the distance travelled by car (Equation~\ref{eq:Car_Simple}).
            According to  \gls{atap} guidelines for road parameter values\footnote{\url{https://www.atap.gov.au/sites/default/files/pv2_road_parameter_values.pdf}} for a medium car with an average journey speed of 60km/h, the estimated fuel coefficient was equal to 11.8 lit/100km.
        	The average annual retail fuel price for the year 2016 in Victoria was \$1.16 according to the Australian Institute of Petroleum data.\footnote{\url{https://aip.com.au/aip-annual-retail-price-data}} 
            Therefore, $\gamma_{d,Car}$ was calculated as:
            
        	\begin{equation}
        	 	\gamma_{d,Car} = \dfrac{11.8 (lit/100km) \times 1.16 (\$) }{100} = 0.137 (\$/km).
        	\end{equation}
        	
            Travel time for each transport mode alternative was another key component of the mode choice model to be estimated. 
            Although the stated travel time for each trip is recorded in \gls{vista}, given they are stated values and not actual, they are often approximations rounded to numbers easier to remember (e.g., quarters, half an hour).
            Furthermore, \gls{vista} only included information for the mode that the traveller chose to use on the survey day, whereas for building a mode choice model, we needed to have travel time for all four alternative travel modes (the one that was chosen as well as those not chosen by the traveller).
            
            We used the Distance Matrix API service\footnote{\url{https://developers.google.com/maps/documentation/distance-matrix/intro}} from the Google Maps platform to estimate travel routes for the final selected \gls{vista} trips and for all transport mode alternatives.\footnote{Google Distance Matrix API is a paid service, not an open data source. However, we made our code to prepare data for the API, sending queries to it and processing its results public in our GitHub repository (\url{https://github.com/matsim-melbourne/calibration-validation}). Additionally, in our script we implemented the option to use OpenRouteService (ORS) instead (\url{https://openrouteservice.org/}), which is open and free to use, with the caveat that ORS does not cover \gls{pt} schedules and congestion.}
    	 	The Google Maps platform was selected as it incorporates congestion and \gls{pt} schedules.
    	 	Therefore, it makes it possible to estimate travel times for different modes based on the current or projected road network, traffic congestion, and \gls{gtfs} schedules.\footnote{The R package \pkg{gmapsdistance (version 3.4)} was used to extract travel times and distances from Google Distance Matrix API.}
    	 	
    	 	One limitation to be considered in this process is that Google uses recent traffic data to estimate travel times. 
    	 	Given the differences in the transport system at the time of using the Google Maps API compared with the \gls{vista} survey day in terms of road infrastructure and traffic behaviour, deviation from the actual time was expected.
    	 	Furthermore, the estimations were extracted from Google Maps API for 09 June 2021,  when Melbourne was still under lockdown due to the COVID19 pandemic outbreak.
        	To account for this deviation, the ``pessimistic'' traffic model from Google Distance Matrix API was used to estimate the travel time for driving.

    	 	Choice model parameters were estimated using \gls{mnl} and based on maximum log-likelihood estimation (MLE).\footnote{The \pkg{mixl} package (version 3.4) in R was used for parameter estimation \citep{molloy_mixl_2019}.} 
    	 	The estimated parameters for the mode choice model of Equation~\ref{eq:SimpleChoiceModel} are presented in Table~\ref{tab:coefficients}.
    	 	These parameters were then used to specify the simulation model's utility function as discussed in the next section.

           \begin{table}
                \begin{center}
                \caption{Estimated mode choice model parameters}
                \label{tab:coefficients}
                \begin{footnotesize}
                \begin{tabular}{l c}
                    \toprule
                    Coefficients                 & Estimation (robse) \\
                    \midrule
                    $\beta_{m}$                   & $0.52^{***}$   $(0.15)$       \\
                    $asc_{PT}$                    & $-1.48^{***}$  $(0.49)$       \\
                    $asc_{Walking}$               & $0.39^{**}$    $(0.17)$       \\
                    $asc_{Cycling}$               & $-3.03^{***}$  $(0.19)$       \\
                    $\beta_{trav, Driving}$       & $-10.42^{***}$ $(1.11)$       \\
                    $\beta_{trav, PT}$            & $-10.52^{***}$ $(1.70)$       \\
                    $\beta_{trav, Walking}$       & $-10.86^{***}$ $(0.92)$       \\
                    $\beta_{trav, Cycling}$       & $-12.56^{***}$ $(1.90)$       \\
                    \midrule
                    \# estimated parameters       & $8.00$         \\
                    Number of respondents         & $14959.00$     \\
                    Number of choice observations & $14959.00$     \\
                    LL(null)                      & $-17252.26$    \\
                    LL(final)                     & $-3758.53$     \\
                    LL(choicemodel)               & $0.00$         \\
                    McFadden R2                   & $0.78$         \\
                    AIC                           & $7533.05$      \\
                    AICc                          & $7533.06$      \\
                    BIC                           & $7593.96$      \\
                    \bottomrule
                    \multicolumn{2}{l}{\tiny{$^{***}p<0.01$; $^{**}p<0.05$; $^{*}p<0.1$}}
                \end{tabular}
                \end{footnotesize}
                \end{center}
            \end{table}
        
    \subsection{Agent-based traffic simulation}\label{method:model}

        The simulation model was based on \gls{matsim} version 13.0 and the inputs from the previous steps.
        The link flow capacity of all network links (Section~\ref{method:network}) was adjusted by a multiplier factor of 0.1 to be compatible with a 10\%  synthetic population sample constructed using the synthetic population generation algorithm (Section~\ref{method:demand}).
        
        Driving, \gls{pt}, cycling, and walking are the four travel modes included in this paper.
        Driving, cycling, and walking were explicitly modelled on the road network, meaning travellers using these modes utilised the road network dedicated to them and the traffic dynamics at each road segment (i.e., a network link) were determined by the \gls{matsim} queue-based road traffic simulator.
        We used the enhanced First-In-First-Out queue model proposed in \citet{agarwal_elegant_2015}, where faster vehicles can pass slower vehicles. 
        Walking and cycling were set to not to block cars in the queue model.
        \gls{pt} vehicle movements were simulated using the deterministic Public Transport Simulation (detPTSim) engine proposed by \citet{metrailler_adding_2018}.
        In detPTSim, \gls{pt} vehicles operate following a strict transit schedule disregarding the queue network and road congestion.
        The use of detPTSim results in a more realistic representation of railway transport (e.g., trains), with potential drawbacks for \gls{pt} vehicles using shared infrastructure with cars (e.g., buses).
                
        The estimated mode choice parameters from Table~\ref{tab:coefficients} were used to construct the \gls{matsim} utility function.
        Specifically, following \citet{horni_multi-agent_2016}, the marginal utility of performing the activity was set to be equal to the marginal utility of travel time by car, and the marginal utility of waiting for \gls{pt} and late arrival were set to be twice and triple this amount, respectively.
        The marginal utility of travel time by car was set to zero, and the marginal utilities of travel time for other modes were adjusted accordingly.
        The resulting values are as listed in Table~\ref{tab:modelParams}.
        These values were used for the initial simulation run, however, as explained later in Section~\ref{sec:outputAnalysis}, mode specific constant values were further calibrated through a number of experiments to improve how well the simulated mode share matched real-world expected values.
        
        \begin{table}[]
        \centering
        \caption{Simulation model utility function parameters}
        \label{tab:modelParams}
            \begin{footnotesize}
                \begin{tabular}{@{}llllll@{}}
                \toprule
                \multicolumn{2}{c}{Model parameters}        & \multicolumn{4}{c}{Value}             \\ \midrule
                \multicolumn{2}{l}{Generic parameters}                       &          &        &         &         \\ 
                 & Marginal utility of money                                 & \multicolumn{4}{c}{0.5159}            \\
                 & Marginal utility of performing activity                   & \multicolumn{4}{c}{10.424}            \\
                 & Marginal utility of late arrival                          & \multicolumn{4}{c}{-31.272}           \\ \midrule 
                \multicolumn{2}{l}{Mode specific parameters}                 & Driving  & PT     & Walking & Cycling \\ \cmidrule(l){3-6} 
                 & Alternative (mode) specific constant                      & 0.0      & -1.483 & 0.385   & -3.033  \\
                 & Marginal Utility of time spent travelling (per hour)      & 0.0      & -0.095 & -0.434  & -2.137  \\
                 & Monetary distance rate (AUD/km)                           & -7.08e-4 & -      & -       & -       \\
                 & Daily monetary cost of using PT (AUD/Day)                 & -        & -8.6   & -       & -       \\
                 & Marginal Utility of waiting at PT station                 & -        & -20.85 & -       & -       \\
                \bottomrule
                \end{tabular}
            \end{footnotesize}
        \end{table}
    
        Both the \textit{workers} and \textit{non-workers} sub-population groups had the \gls{matsim} innovation strategy for route choice (\textit{re-routing strategy}) enabled for their re-planning step   (Figure~\ref{fig:matsimLoop}).
        The \textit{sub-tour mode choice strategy} was also enabled for the workers sub-population, which allowed them to change their trip leg modes and to find the one that works best for them.
        All four main modes (i.e., driving, \gls{pt}, walking, and cycling) were available for all worker agents and cycling and driving were set to be a \textit{tour-mode} meaning that for an agent to have driving in one of its trip legs it must start the trip tour from home with a car and must return to home with a car as well.
        No innovation strategy for activity type, location, or timing selection was included. 
        These attributes were considered constant during the simulation since they were generated by and calibrated in the  synthetic population generation process.
        
        If in the re-planning step of a simulation iteration neither of these two innovation strategies were adopted by an agent, the agent was set to change its plan to another previously experienced plan from its memory (memory size = 5 highest scored experienced plans) with probability $e^{\Delta Score}$, where $\Delta Score$ is the difference in scores between the two plans.
        This strategy, known as \textit{ChangeExpBeta} strategy, was selected to encourage agents to seek plans that yield globally optimal scores. 
        More in-depth discussion about this innovation strategy can be found in \citet{nagel_agent-based_2016}.
      	The weighting of each of the innovation strategies for each subgroup is listed in Table~\ref{tab:strategies}.
        
        The simulation model was run for 200 iterations and re-routing and sub-tour mode choice innovation strategies were disabled for the last 40 iterations (20\%) to allow the model to converge to a stable solution (net score).
        
        \begin{table}[]
            \caption{Re-planning innovation strategy weights for different sub-populations}
            \centering
            \label{tab:strategies}
            \begin{tabular}{@{}lcc@{}}
                \toprule
                \multirow{2}{*}{Strategy} & \multicolumn{2}{c}{Weight} \\ \cmidrule(l){2-3} 
                                          & Workers    & Non-workers   \\ \midrule
                ChangeExpBeta             & 0.8        & 0.9           \\
                Re-routing                & 0.1        & 0.1           \\
                Sub-tour mode choice      & 0.1        & 0.0           \\ \bottomrule
            \end{tabular}
        \end{table}
    
        Lastly, the \pkg{SwissRailRaptor} extension to \gls{matsim} was used as the \gls{pt} router \citep{metrailler_adding_2018}. 
        The SwissRailRaptor extension provides a significantly more efficient \gls{pt} routing in \gls{matsim} and adds additional features for more realistic \gls{pt} simulation.
        One of these is the inter-modal access and egress feature that we used for modelling trips to/from \gls{pt} stops as described below.
        In this paper, walking was the only travel mode considered for access/egress trip legs.
        Potential start and end stops for each \gls{pt} trip leg were filtered to those within a certain radius of the trip leg's origin and destination.  
        The initial value of this search radius was set to 1km. 
        If fewer than two stops were found in this radius, it was increased by another 1km until at least two stops are found or a maximum radius of 10km is reached.
        It should be noted that the search radius of 1km does not mean that agents travel 1km to get to their desired \gls{pt} stop, only that agents consider all stops within this radius as their potential candidates, and will select the best one based on various factors including the amount of walking they have to do and the transit lines servicing each stop.
        
\section{Simulation output analysis}\label{sec:outputAnalysis}
    
    This section analyses and compares the simulation outputs with the real-world observations to better understand the accuracy and reliability of our model.
    To achieve this, three main measures of mode share (Section~\ref{method:modeShareComp}), road traffic volume (Section~\ref{method:roadTraffic}), and travel distance and time (Section~\ref{sec:distTimeAnalysis}) were analysed.
    
    \subsection{Mode Share analysis}\label{method:modeShareComp}
    
        As explained in Section~\ref{method:model}, \gls{matsim} sub-tour mode choice strategy was enabled for the \textit{workers} sub-population.
        To examine and calibrate the mode choice model for this sub-population, we compared simulated trips to work with the \gls{vista} 2016-18 survey data commute to work trips \citep{department_of_transport_victorian_2018}, as well as \gls{abs} Census 2016 Method of Travel to Work (MTW) data for Greater Melbourne \citep{australian_bureau_of_statistics_abs_method_2012}.
        Census MTW data was accessed through ABS TableBuilder Pro online tool \citep{australian_bureau_of_statistics_abs_tablebuilder_2016} and was filtered to include only the four travel modes included in this paper (i.e., driving, \gls{pt}, walking, and cycling). 
        
        We then followed an iterative process for manual calibration of the mode choice functionality of the model.
        First, we ran the simulation model with the parameters listed in Table~\ref{tab:modelParams} for 200 iterations, allowing agents to find their best travel mode and route given these estimated parameters.
        The mode share of the simulation output was then compared to the expected real-world values from Census MTW data, and the model's mode-specific constants were adjusted to achieve a better match.
        Next, we ran another simulation experiment for 100 iterations with the new adjusted parameters and using the already optimised plans from the previous run.
        This iterative process of adjusting parameters, running the simulation, and comparing the results mode shares with Census MTW 2016 was repeated until a reasonable fit was achieved.
        We considered the mode shares from the simulation output for trips to work to be within $\pm1\%$ error threshold of the observation from Census MTW as our calibration target.
        The final calibrated simulation mode shares for trips to work and the adjusted value of the mode-specific constants are listed in Table~\ref{tab:modeShareComp} and Table~\ref{tab:adjustedConstants}, respectively.
        
        \begin{table}[h]
            \centering
            \caption{Adjusted mode specific constants as a result of the mode share calibration}
            \label{tab:adjustedConstants}
            \begin{footnotesize}
            \begin{tabular}{@{}lllll@{}}                                                    \toprule
                                                & Driving  & PT     & Walking & Cycling \\  \midrule
            Adjusted mode specific constants    & 0.0      & -1.483 & 0.385   & -3.033 \\   \bottomrule
            \end{tabular}
            \end{footnotesize}
        \end{table}
        
        The share of non-work trips was also compared with real-world data to examine if enabling the sub-tour mode choice strategy for the workers sub-population was acceptable or a mode choice strategy for both sub-population groups was needed.
        To do this, the mode share in all non-work trips from the calibrated simulation output was compared to the share of these travel modes in \gls{vista} 2016-18 non-work trips.
        Table~\ref{tab:modeShareComp} provides mode shares for the mode choice calibrated simulation model, VISTA travel survey data 2016-18, and Census MTW 2016 for work and non-work trips. 
        
        \begin{table}[ht]
            \caption{Mode share comparison between calibrated simulation output, VISTA 2016-18 and Census MTW 2016}
            \label{tab:modeShareComp}
            \centering
            \small{
            \begin{tabular}{rllccc}
              \toprule
              \multicolumn{3}{l}{} & \multicolumn{3}{c}{Mode share (\%)} \\ \cmidrule{4-6}
              &&& Simulation & VISTA 2016-18 & Census MTW 2016 \\
              \midrule
                \multicolumn{5}{l}{Trips to work} \\ 
                  &     & Driving   & 74.8  & 73.4  & 75.2\\ 
                  &     & PT        & 21.5  & 21.4  & 19.3\\ 
                  &     & Walking   & 2.1   & 2.5   & 3.7 \\ 
                  &     & Cycling   & 1.6   & 2.7   & 1.7 \\ \midrule
                \multicolumn{5}{l}{Non work trips} \\ 
                  &     & Driving   & 70.2  & 64.1  &  - \\ 
                  &     & PT        & 13.2  & 10.2   &  - \\ 
                  &     & Walking   & 15.4  & 23.1  & -  \\ 
                  &     & Cycling   & 1.2   & 2.6   & -  \\ 
               \bottomrule
            \end{tabular}
            }
        \end{table}

    \subsection{Road traffic volume analysis}\label{method:roadTraffic}
        
        \begin{figure}[h]
        	\centering
        	\begin{subfigure}[b]{0.45\textwidth}
                \includegraphics[trim={2cm 2cm 1.5cm 1.5cm},clip,width=\textwidth]{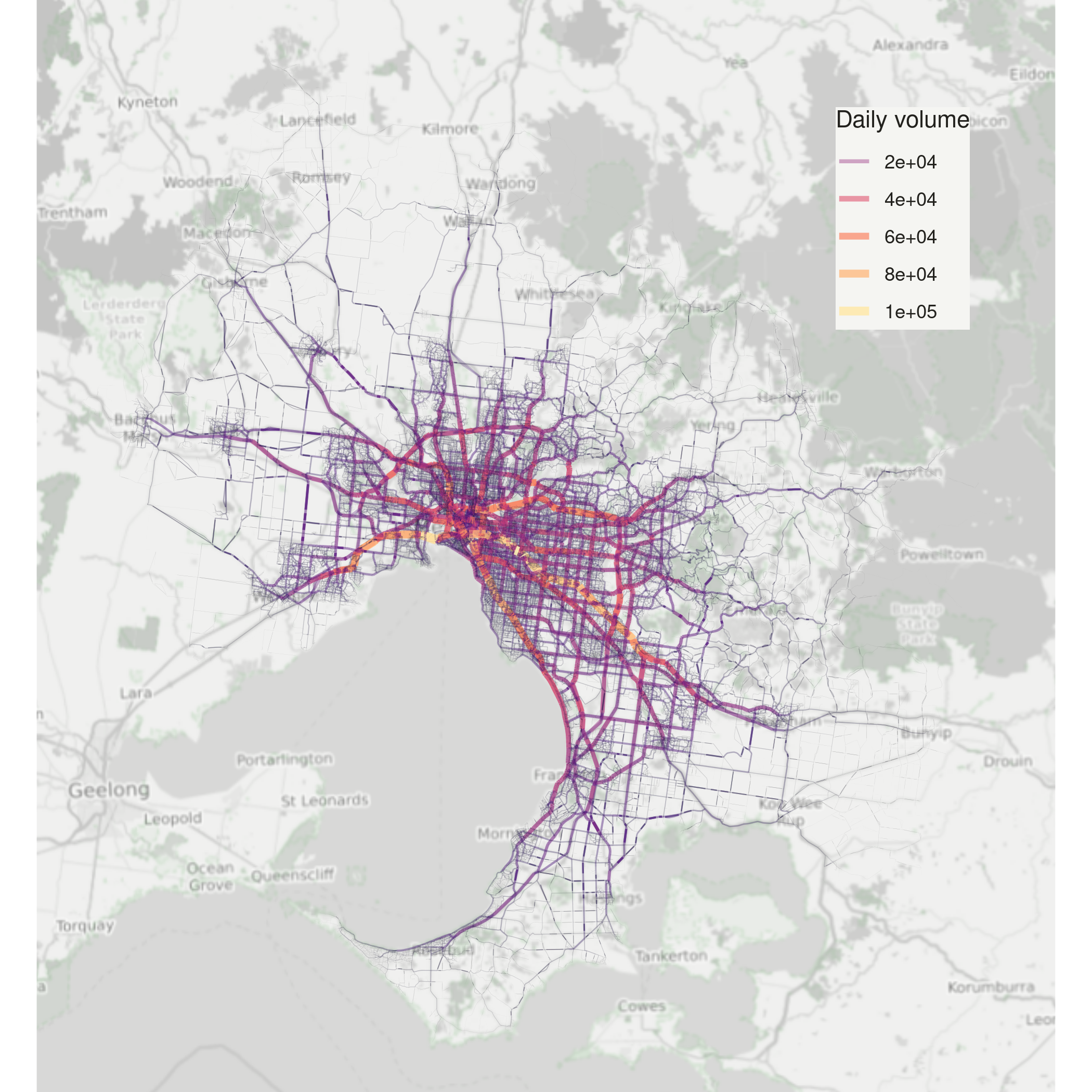}
            	\caption{Driving}
                \label{fig:car_daily}
            \end{subfigure}
        	\begin{subfigure}[b]{0.45\textwidth}
                \includegraphics[trim={2cm 2cm 1.5cm 1.5cm},clip,width=\textwidth]{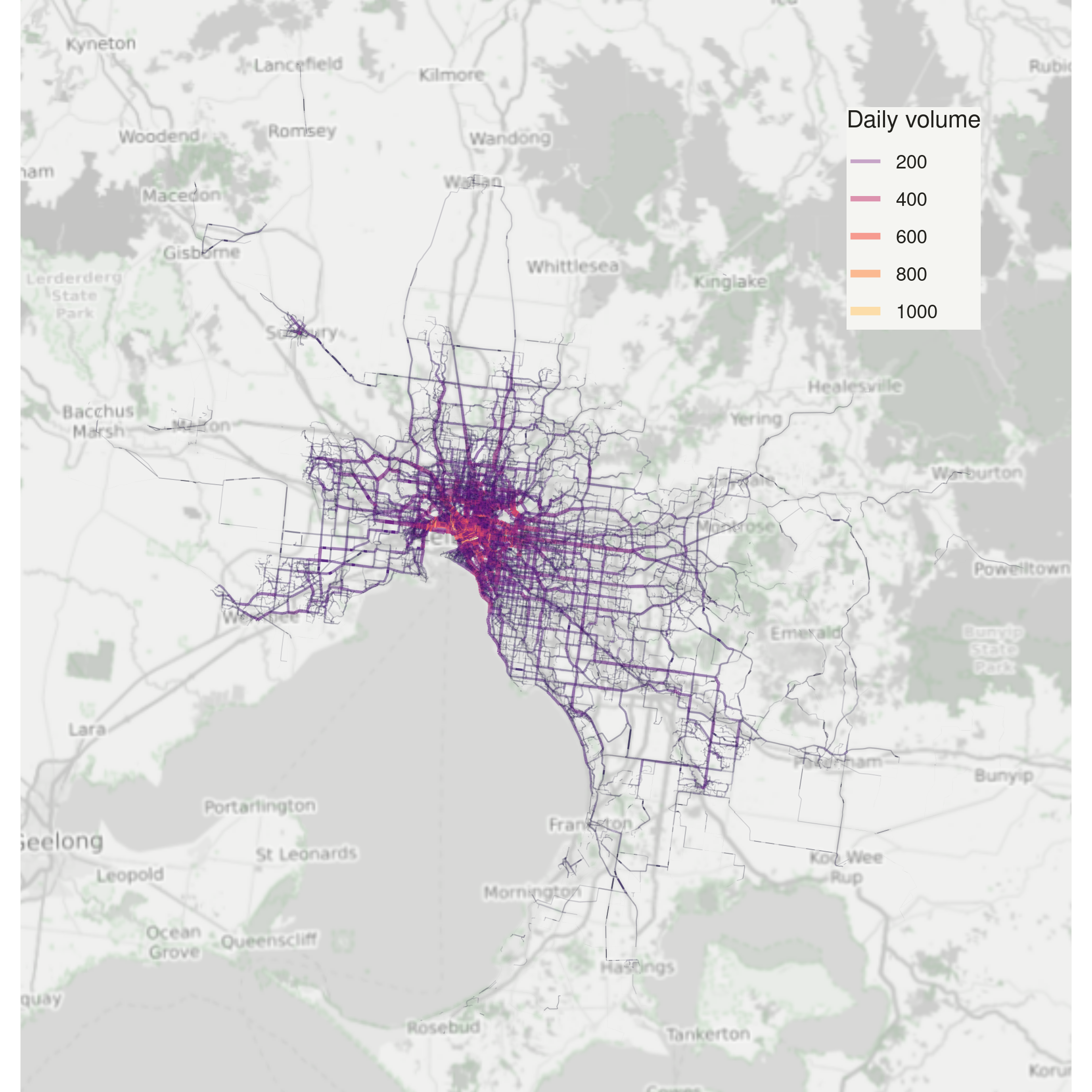}
            	\caption{Cycling}
                \label{fig:bike_daily}
            \end{subfigure}
        	\begin{subfigure}[b]{0.45\textwidth}
                \includegraphics[trim={2cm 2cm 1.5cm 1.5cm},clip,width=\textwidth]{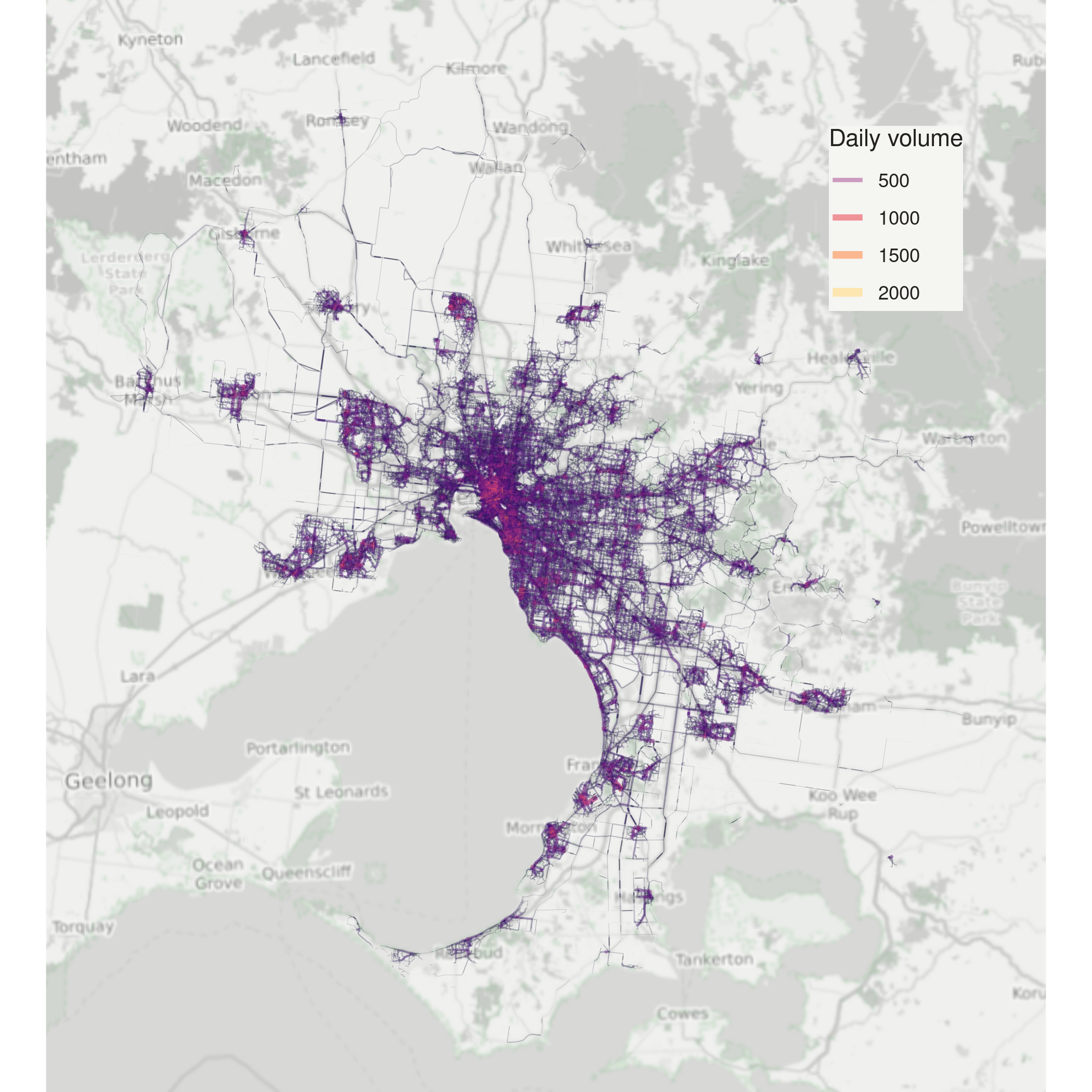}
            	\caption{Walking}
                \label{fig:walk_daily}
            \end{subfigure}
        	\caption{Simulation output aggregated daily traffic volume for different modes}
        	\label{fig:dailyTrafficVols}
        \end{figure}
        
        Daily driving, cycling, and walking traffic volumes from the calibrated simulation output are illustrated in Figure~\ref{fig:dailyTrafficVols}.
        
        Publicly available traffic count data for Melbourne was used to examine the road usage accuracy of the model for driving.
        We used the \gls{thtv} data from Victoria’s open data platform for 2019,\footnote{\url{https://data.vicroads.vic.gov.au/Metadata/Typical\%20Hourly\%20Traffic\%20Volumes.html}, accessed on 14/05/2021} which provides the typical traffic volumes for major arterial roads across Victoria.
        \gls{thtv} was filtered down to the data for school term normal mid-week days.
        Then the data were divided into two categories of roads: towards Melbourne CBD (handling most of the AM peak traffic) and those going outward from the CBD (handling PM peak traffic).
        In each category (i.e., AM and PM), the top 10\% highest traffic roads were identified, and within each, the road segment with the maximum volume was selected for comparison with the simulation output.
        This resulted in 87 road segments, 47 for AM peak and 40 for PM peak hours, being selected for further analysis.
        
        The selected road segments were joined to their equivalent links in the simulation road network. For this purpose, the 'equivalent' link was selected as the link located closest to the midpoint of the road segment that satisfied the conditions of operating in the correct direction to match the road, and having a bearing (or azimuth) within 17.5 degrees of the bearing of the road segment.
        
        For the cycling traffic volume comparison, the average weekday daily cycling volume from automatic cycling volume and speed sensors for Greater Melbourne was used, downloaded from Victoria’s open data platform for the period of March 2019.\footnote{\url{https://discover.data.vic.gov.au/dataset/bicycle-volume-and-speed}}
        Each sensor  was joined to its equivalent link in the simulation road network, by selecting the closest link that was either a bicycle path or a road with a bicycle lane and that operated in the correct direction.
        In total, 48 counting sensors (some mono-directional and others bi-directional), corresponding to 70 network links, were selected for further analysis.
        
        For walking traffic volume, we used pedestrian counting automated sensor data located across Melbourne's central \gls{lga}, i.e., City of Melbourne, encompassing the Central Business District and surroundings \citep{city_of_melbourne_pedestrian_2021}.
        The data was downloaded for mid-week work days of March 2019 and was joined to the simulation road network by selecting the closest links having similar bearings to the relevant footpaths, in a similar way as described above for driving volumes.
        Given the footpaths are bi-directional, the aggregated number of pedestrians passing each sensor, regardless of the walking direction, were used for comparison.
        Furthermore, for streets with more than one footpath, the aggregated volume from all associated footpaths was used.
        This resulted in selecting 48 sensors corresponding to 93 network links for further analysis.
        
\begin{figure}[h]
        	\centering
        	\begin{subfigure}[b]{0.45\textwidth}
        		\includegraphics[width=\textwidth]{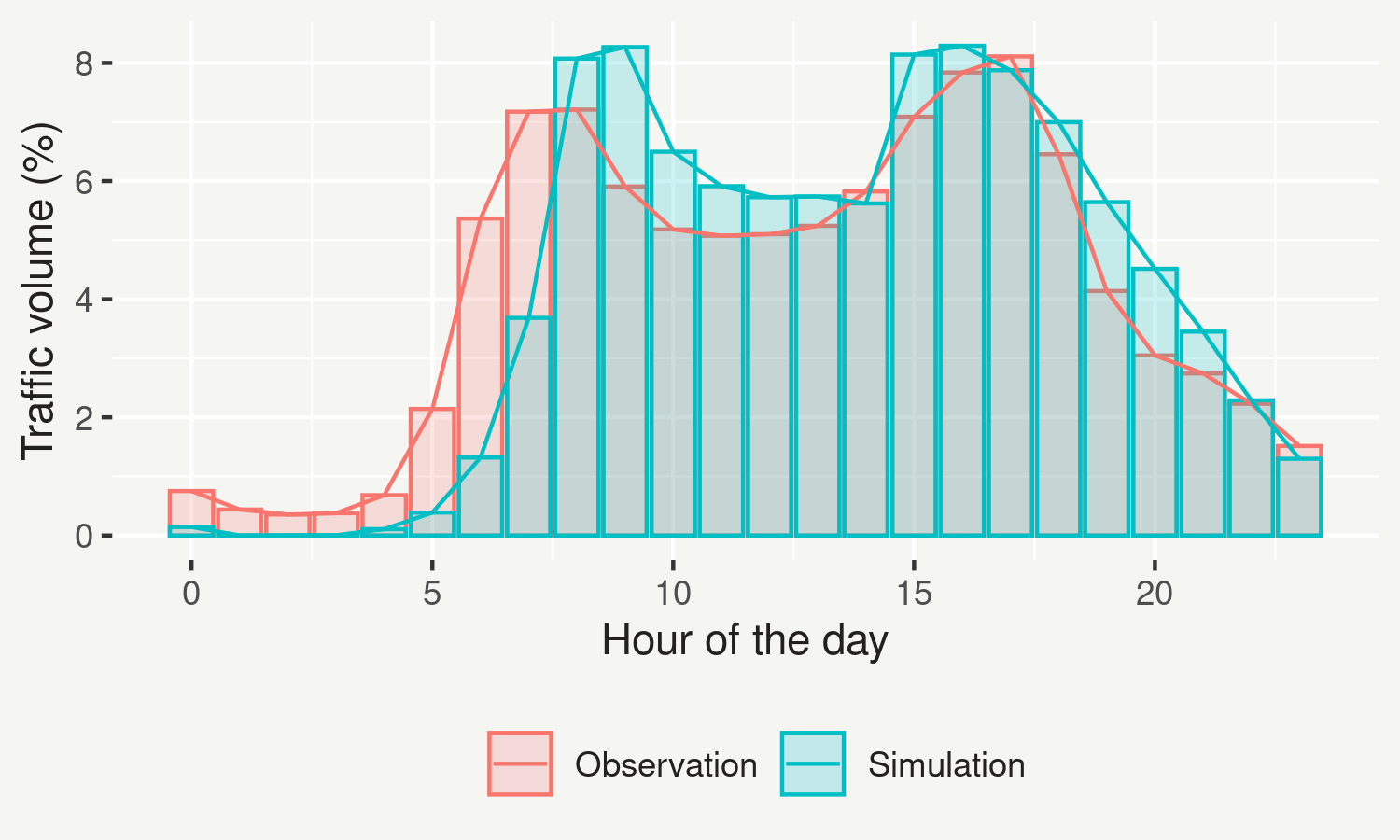} 
        		\caption{Driving}
        		\label{fig:carTotalVols}
        	\end{subfigure}
            \begin{subfigure}[b]{0.45\textwidth}
        		\includegraphics[width=\textwidth]{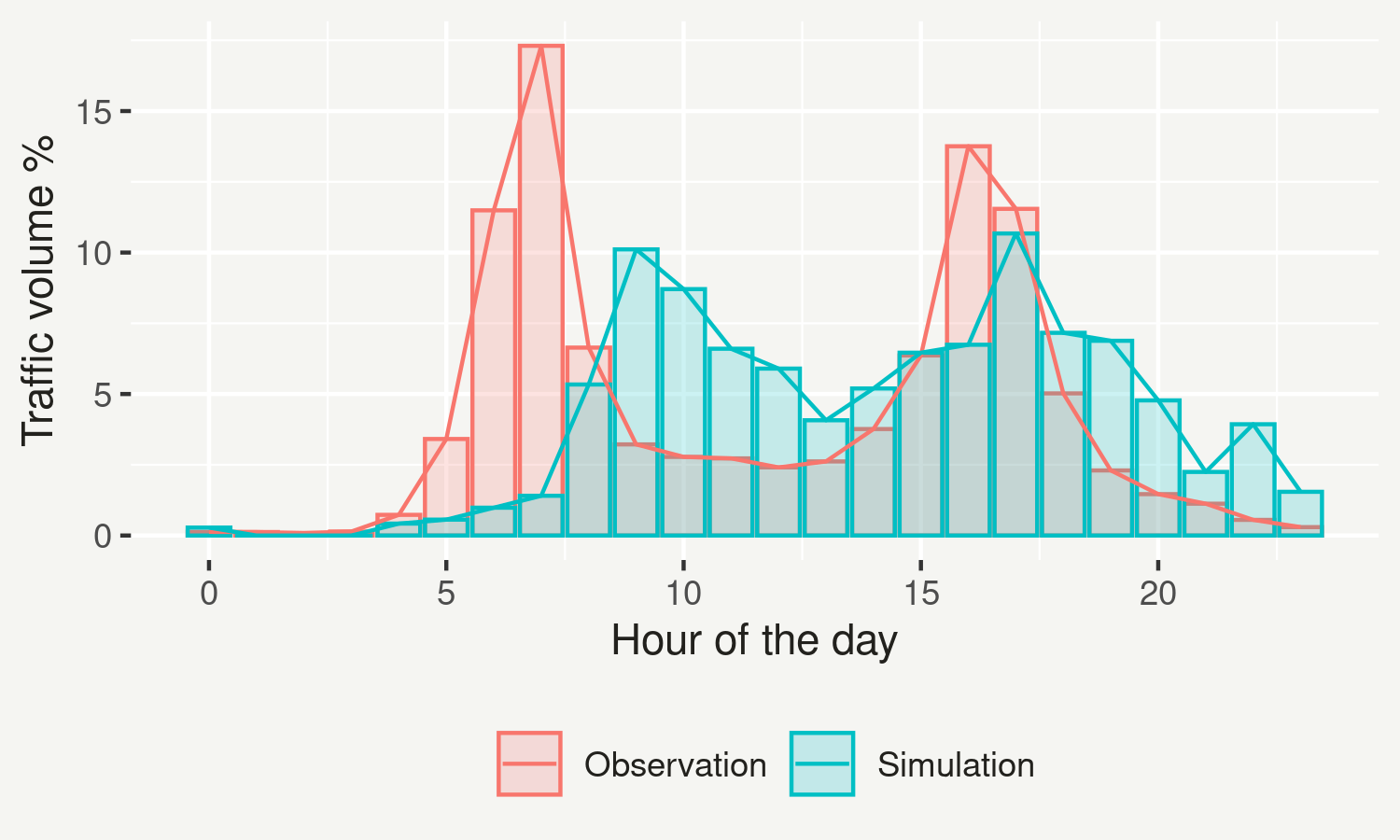} 
        		\caption{Cycling}
        		\label{fig:bikeTotalVols}
        	\end{subfigure}
        	\begin{subfigure}[b]{0.45\textwidth}
        		\includegraphics[width=\textwidth]{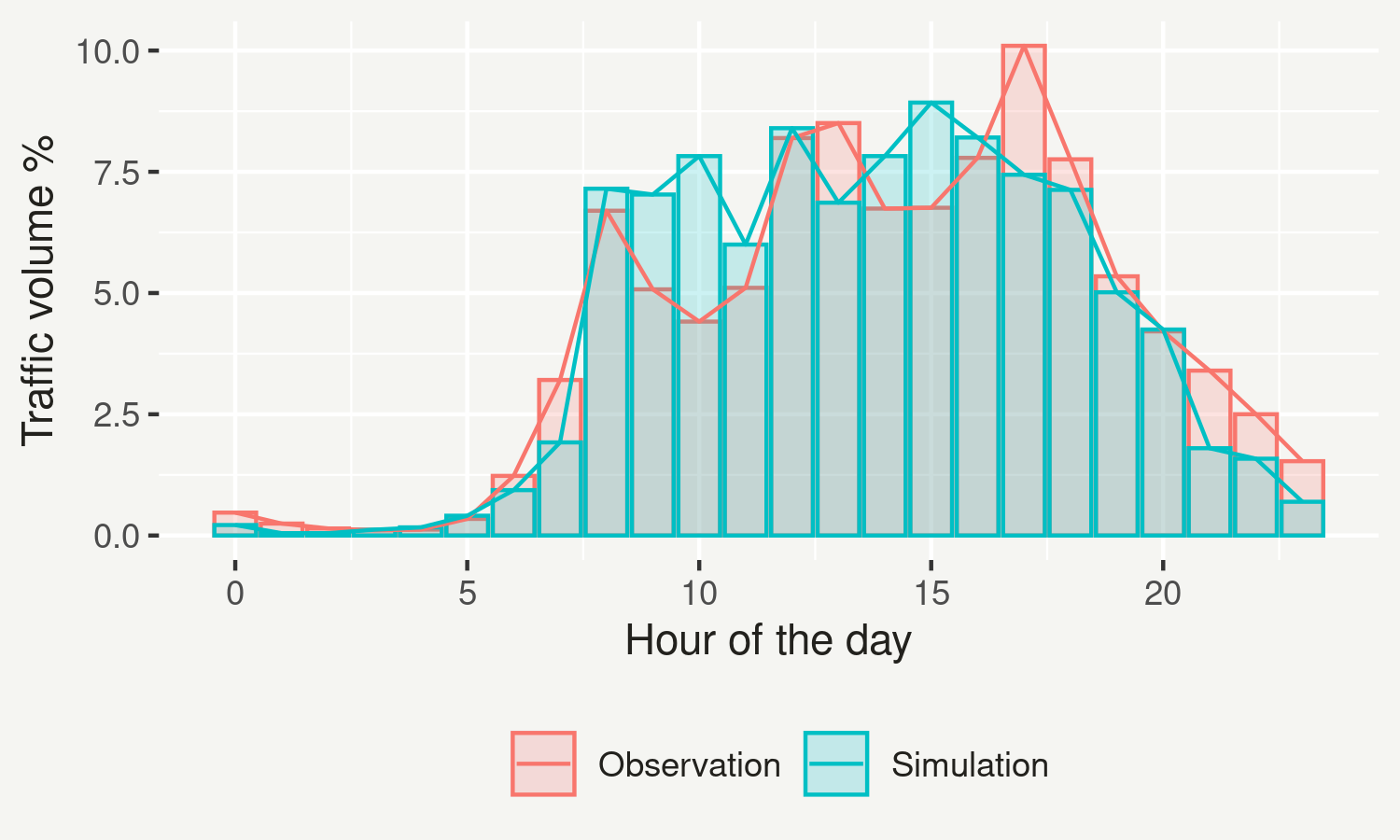}
        		\caption{Walking}
        		\label{fig:walkTotalVols}
        	\end{subfigure}
        	\caption{ Aggregated hourly traffic volume percentages in simulation versus observation for different travel modes.}
        	\label{fig:TotalVolsComp}
        \end{figure}

        The traffic volume percentage of the daily traffic for every hour of the day, $h$, and for every road segment was calculated for the calibrated simulation output, $s'_{h}$ and the observation data from \gls{thtv}, $s_{h}$, using Equation~\ref{eq:volPct}.
        
        \begin{equation}\label{eq:volPct}
            s_{r,h} = \frac{ v_{r,h} }{ \sum_{h=0}^{23} v_{r,h} }, 
        \end{equation}
        
        where $N$ is the total number of road segments analysed, and $v_{r,h}$ is the traffic volume of road $r$ during the hour $h$.
        Figure~\ref{fig:carTotalVols} depicts the average traffic volume percentage of the daily traffic for every hour of the day across all selected $N=87$ road segments. 
        We then used Weighted Absolute Percentage Error (WAPE) to compare the hourly road traffic volume percentages in the observation data and simulation outputs (Equation~\ref{eq:wape}).
        
        \begin{equation}\label{eq:wape}
            WAPE_{h} = \frac{\sum_{r=1}^N \lvert s_{r,h}-s'_{r,h} \rvert}{\sum_{r=1}^N \lvert s_{r,h} \rvert}.
        \end{equation}
        
\begin{figure}[h]
        	\centering
        	\begin{subfigure}[b]{0.45\textwidth}
        		\includegraphics[width=\textwidth]{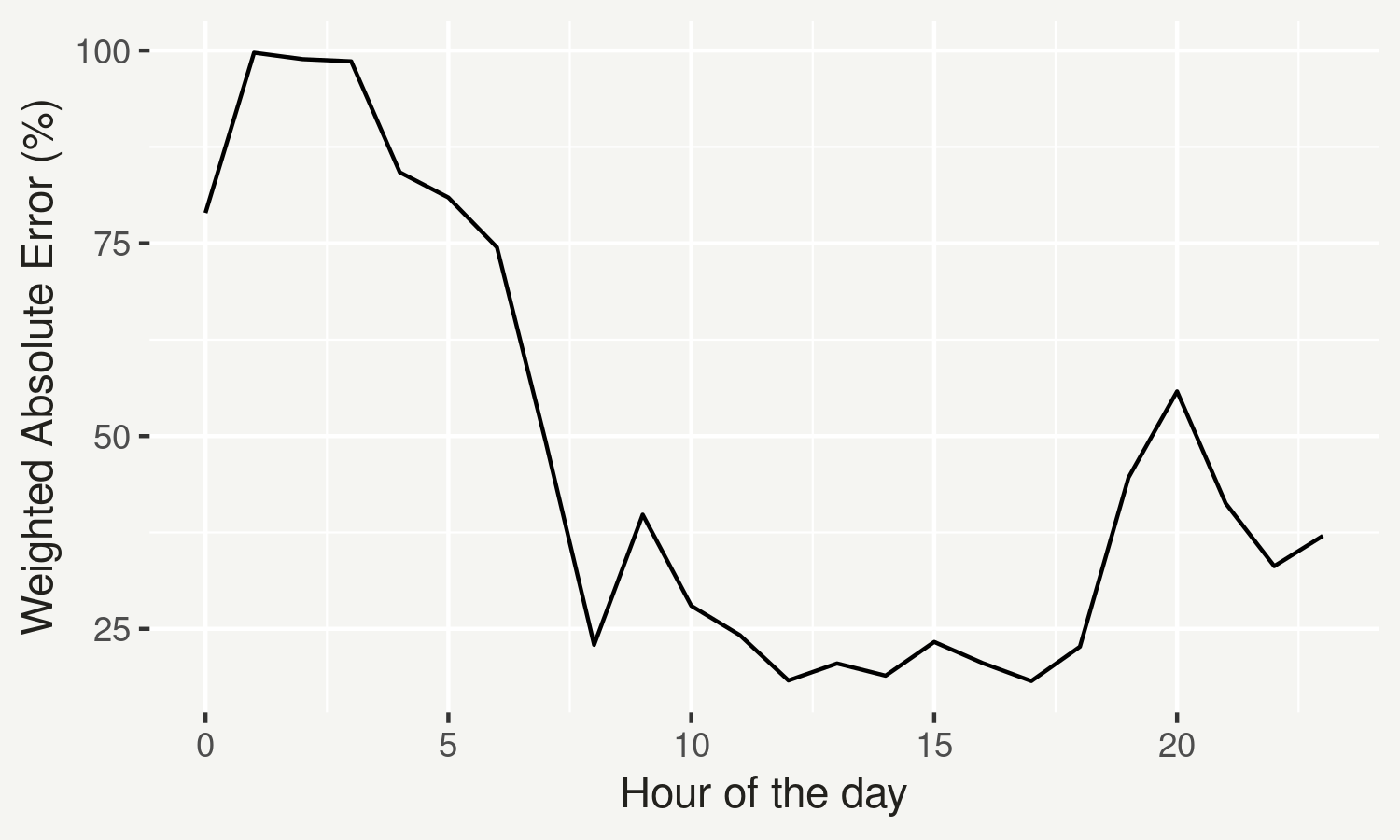} 
        		\caption{Driving}
        		\label{fig:carTotalWape}
        	\end{subfigure}
            \begin{subfigure}[b]{0.45\textwidth}
        		\includegraphics[width=\textwidth]{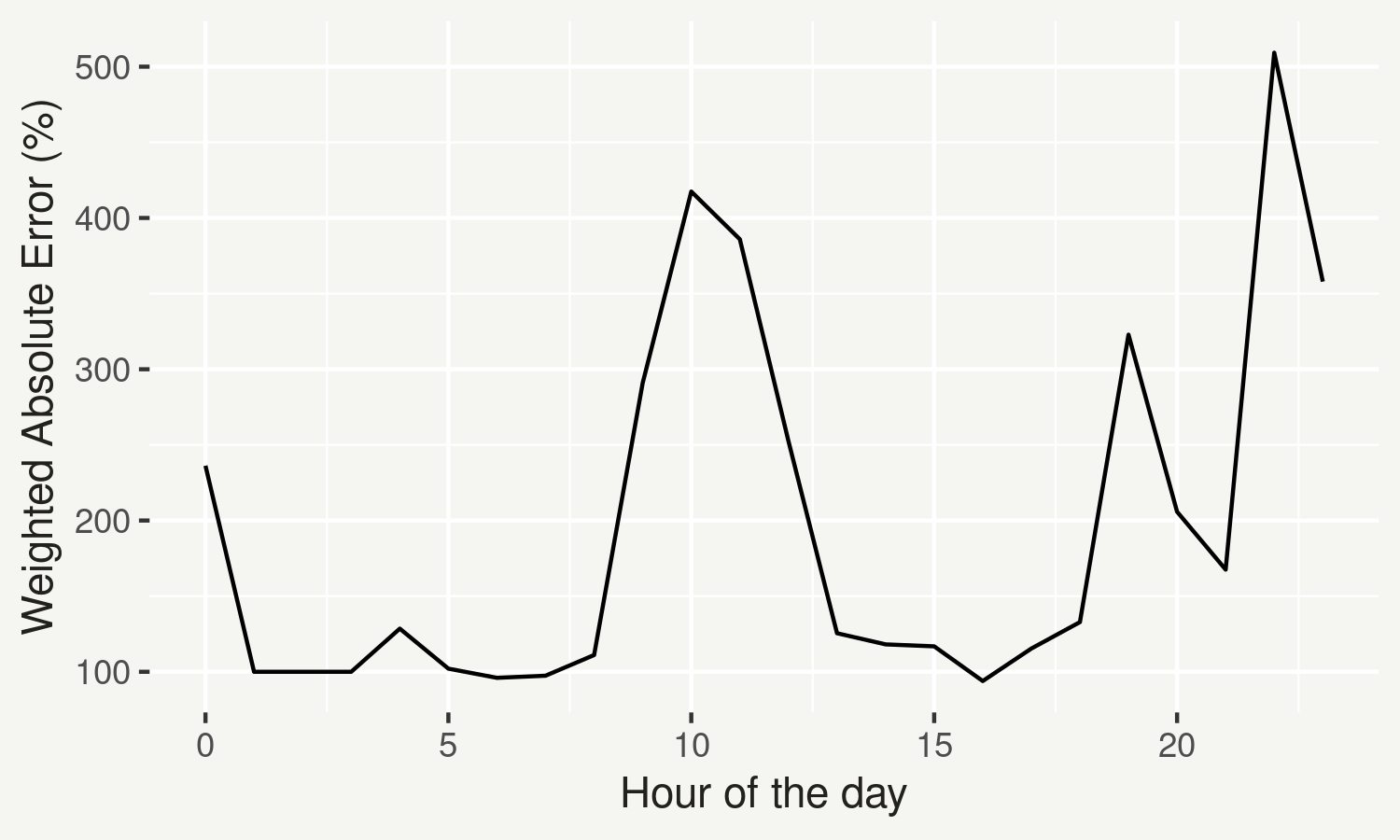} 
        		\caption{Cycling}
        		\label{fig:bikeTotalWape}
        	\end{subfigure}
        	\begin{subfigure}[b]{0.45\textwidth}
        		\includegraphics[width=\textwidth]{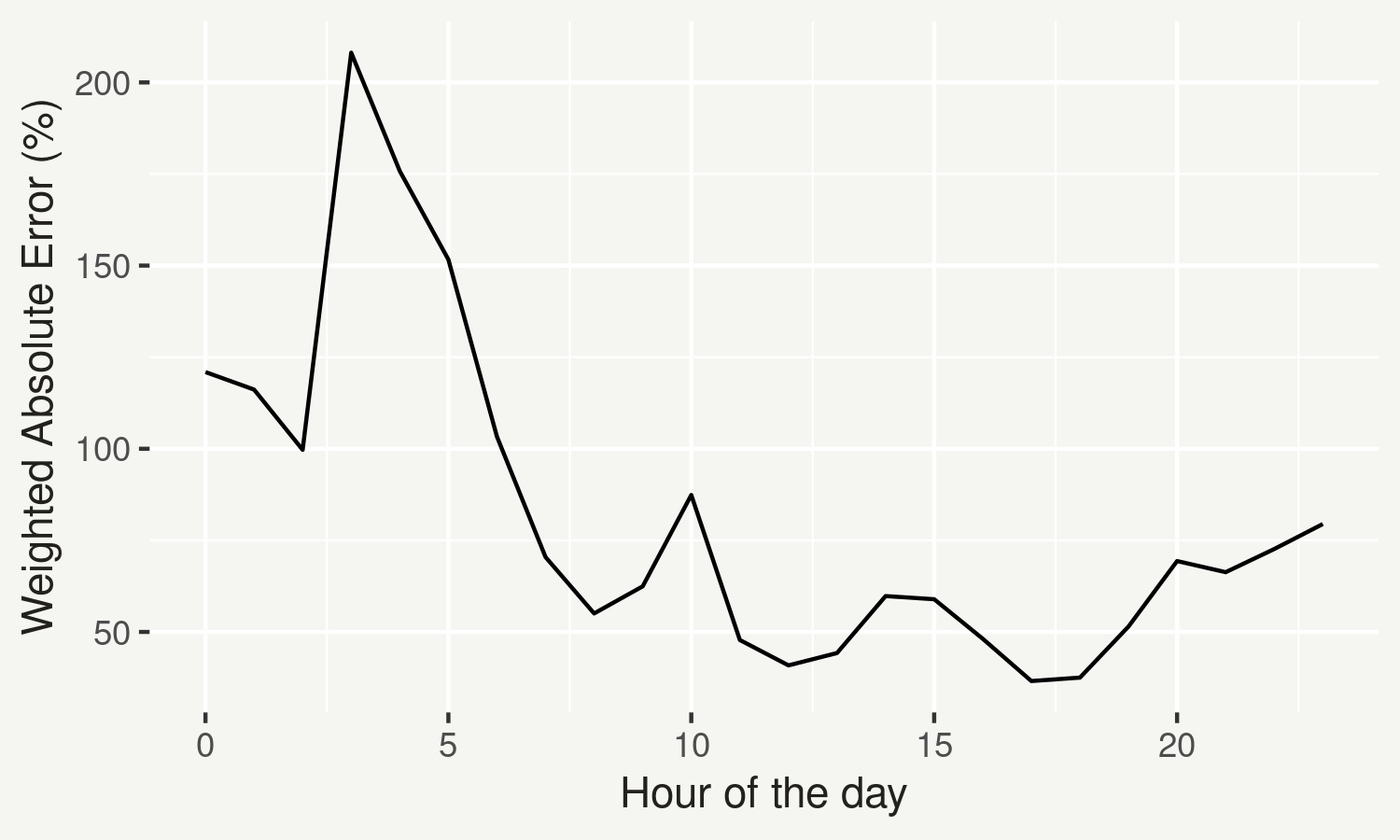}
        		\caption{Walking}
        		\label{fig:walkTotalWape}
        	\end{subfigure}
        	\caption{Weighted absolute percentage error of aggregated hourly traffic volume percentages in simulation versus observation for different travel modes.}
        	\label{fig:TotalWape}
        \end{figure}

        As shown in Figures~\ref{fig:carTotalWape} and \ref{fig:carTotalVols} the simulation model does well in capturing the peak hours car traffic volume in Melbourne with WAPE under 25\%.
        A potential reason for the car traffic volume deviations in the early morning and late evening was not having freight traffic, travellers from outside the Greater Melbourne area and airport passengers in the current version of the model, whose absence is likely to be more noticeable during off-peak hours when roads are not already congested with local commuters.
        A similar trend was also observed for walking as shown in Figure~\ref{fig:walkTotalWape}.
        For cycling, however, the traffic volume percentage error was high throughout the day, and considerably higher at off-peak hours (Figures~\ref{fig:bikeTotalWape} and \ref{fig:TotalVolsComp}).
        This was likely due to not including the impact of cycling-relevant road infrastructure, such as bikeway type or slope, on cycling route choice behaviour as discussed further in Section~\ref{sec:discussion}.
        
    \subsection{Public transport usage analysis}
        
\begin{figure}
            \centering
            \includegraphics[width=0.8\textwidth]{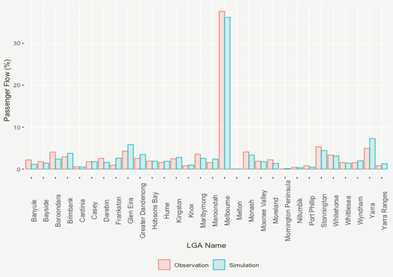}
            \caption{Passenger flow percentage comparison at the LGA level in real-world and simulation outputs.}
            \label{fig:ptflow}
        \end{figure}

        To validate the public transport usage in the model, the real-world percentage of passenger flow aggregated at \gls{lga} level was compared with the outputs of our simulation. 
        To calculate this percentage, the passenger flow of 218 train stations across the Greater Melbourne Metropolitan area was aggregated based on the \gls{lga} they were located within. 
        Then, the aggregated share of the passenger flow of each \gls{lga} relative to the total passengers of Greater Melbourne was calculated for further comparison. 
        Station Access survey data from Public Transport Victoria for 2016 was used for this comparison. 
        This data was obtained from Victoria's Department of Transport -- Public Transport Victoria.
        Figure~\ref{fig:ptflow} shows the comparison of real-world observations and our simulation outputs, indicating that the simulation model was able to capture the \gls{pt} passenger flow distribution across the Greater Melbourne.
        
    \subsection{Travel distance and time analysis}\label{sec:distTimeAnalysis}
        
        Mode share and road traffic volume analyses evaluated the model at an aggregated level, i.e., either aggregated to the travel modes or road segments.
        We conducted another analysis on travel distance and time of a sample of trips to also evaluate the model at the individual trip level.
        To do this, a subset of 1,000 trips, stratified by the origin \gls{sa3} and travel mode, were randomly sampled from the simulated trips.
        Experienced travel distance and time for the sample trips were extracted from the simulation output.
    
        The simulated travel time for driving incorporated the impact of road congestion in addition to speed limits and the vehicle's maximum speed.
        This is due to the \gls{matsim} queue model for capturing road traffic for the travel modes being set to use the road network. 
        Walking and cycling were also network modes, however, they were set not to impact or be impacted by the road traffic, hence their travel time and distance were simplified reflective of the network distance between origin and destination and their constant speeds, $1.7\nicefrac{m}{s}$ for walking and $5.5\nicefrac{m}{s}$ for cycling. 
        \gls{pt} travel time was based on the transit schedules extracted from \gls{gtfs} and the agent's decision about which \gls{pt} service to use.
        
        We used the Google Distance API to estimate the expected travel distance and time for the sampled trips.
        For driving and \gls{pt}, Google Distance API estimates travel distance and time based on its historical records, taking into consideration the traffic and network conditions.
        For cycling and walking, Google only assumes the fastest route.
        Another limitation of using Google Distance API is that it does not provide estimates for a past trip. 
        Therefore, the travel distances and time of a sample of trips were estimated based on October 2021 Google data.
        
\begin{figure}[h]
        \centering
        \begin{subfigure}[b]{0.4\textwidth}
        	\includegraphics[width=\textwidth]{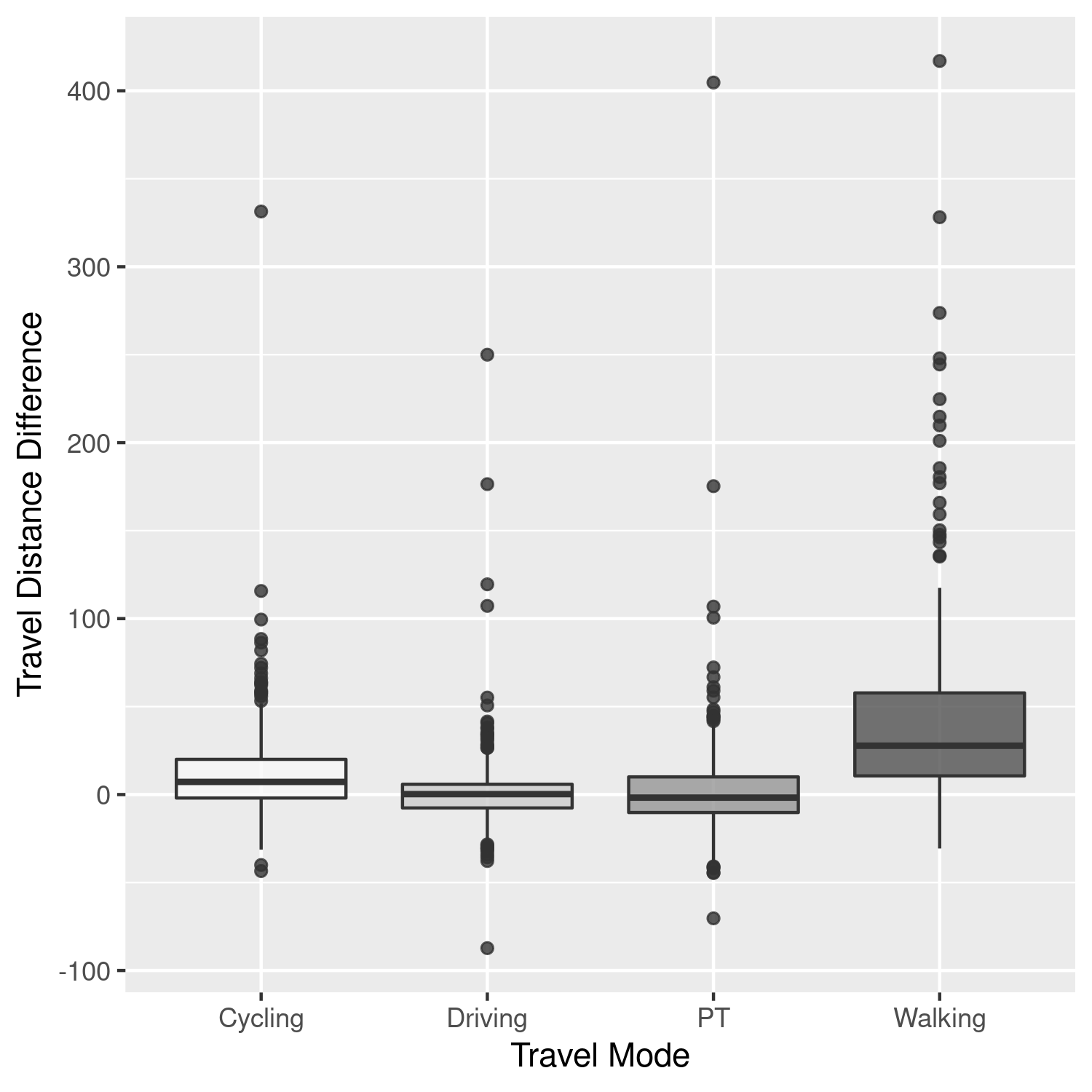} 
        	\caption{Travel distance (\%)}
        	\label{fig:travDistDiff}
        \end{subfigure}
        \begin{subfigure}[b]{0.4\textwidth}
        	\includegraphics[width=\textwidth]{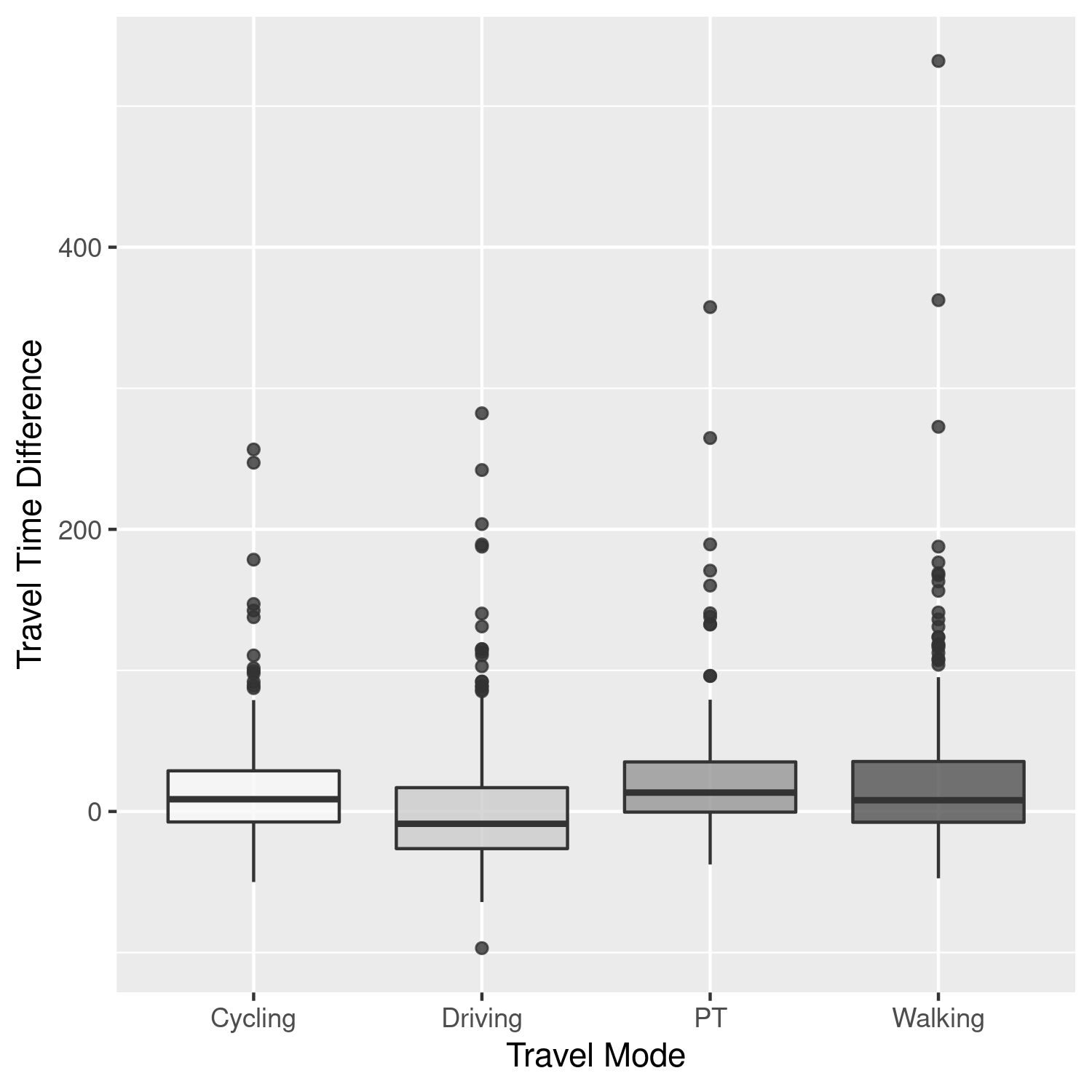} 
        	\caption{Travel time (\%)}
        	\label{fig:travTimeDiff}
        \end{subfigure}
        \caption{Percentage error of travel (a) distance and (b) time between simulation output and Google Distance Matrix API estimates for sampled trips}
        \label{fig:timeDistDiff}
        \end{figure}

         Figure~\ref{fig:timeDistDiff} illustrates the percentage error of travel distance and travel time for different modes for the 1,000 sampled trips.
         For example, the percentage error of travel distance for a sample trip $j$ and mode $m$, $\epsilon_{m,j}^{d}$, was calculated as follows:
        
        \begin{equation}\label{eq:distDiff}
            \epsilon_{m,j}^{d} = 100 \times \frac{d'_{m,j} - d_{m,j}}{d_{m,j}} ,
        \end{equation}
        where $d'_{m,j}$ is the experienced travel distance from the simulation for mode $m$ and trip $j$ and $d_{m,j}$ is the expected travel distance from Google Distance API.
        
\section{Discussion and Conclusion}\label{sec:discussion}

    In this paper, we developed an open\footnote{For the steps that closed data were used for better accuracy or due to the availability of data, we made output estimations or rasterised information extracted from them  openly available so that the complete workflow can be reproduced by the user community.} multi-modal activity-based and agent-based model for the Greater Melbourne area.
    We described the complete workflow of the model development from creating the simulation scenario inputs (i.e., road network, synthetic population, and mode choice parameters), to mode choice model calibration and simulation output analysis.
    All the tools we described and developed for the \gls{atom} model are open and publicly available in our GitHub repository.
    Furthermore, these tools were designed to utilise data sources that are commonly available for different cities around the world (e.g., travel surveys, traffic counts, \gls{osm}, and \gls{gtfs}).
    This means although the format of some of the data used in this paper might be specific to Melbourne, such as VISTA or traffic counts, the same workflow could be used for other cities if the data structure is compatible with the expected structure of each tool. 
    
    The tools we presented as part of the model development workflow, such as population demand,\footnote{\url{https://github.com/orgs/matsim-melbourne/demand}} network supply generation,\footnote{\url{https://github.com/orgs/matsim-melbourne/network}} and mode choice model  estimation\footnote{\url{https://github.com/matsim-melbourne/choice-model}} processes, serve as standalone models in their own right.
    Therefore, these models are suitable for more general use outside of \gls{matsim}. 
    
    We calibrated the mode choice behaviour of the work trips for four travel modes of driving, \gls{pt}, walking, and cycling against the ABS Census Method of Travel to Work 2016 and also \gls{vista} 2016-18 (Table~\ref{tab:modeShareComp}). 
    The simulated mode share percentage for non-work trips also resembled the figures observed in the travel survey. 
    The mode choice calibrated simulation model could be used as the baseline for examining the potential for mode shift as a result of the built environment, infrastructure and/or monetary interventions, such as constructing new roads or modifying an existing road, increasing \gls{pt} services to existing stations, adding new stations and service lines, or changing \gls{pt} fares or motor vehicle fuel prices.
    
\begin{figure}[h]
        \centering
        \begin{subfigure}[b]{0.47\textwidth}
        	\includegraphics[width=\textwidth]{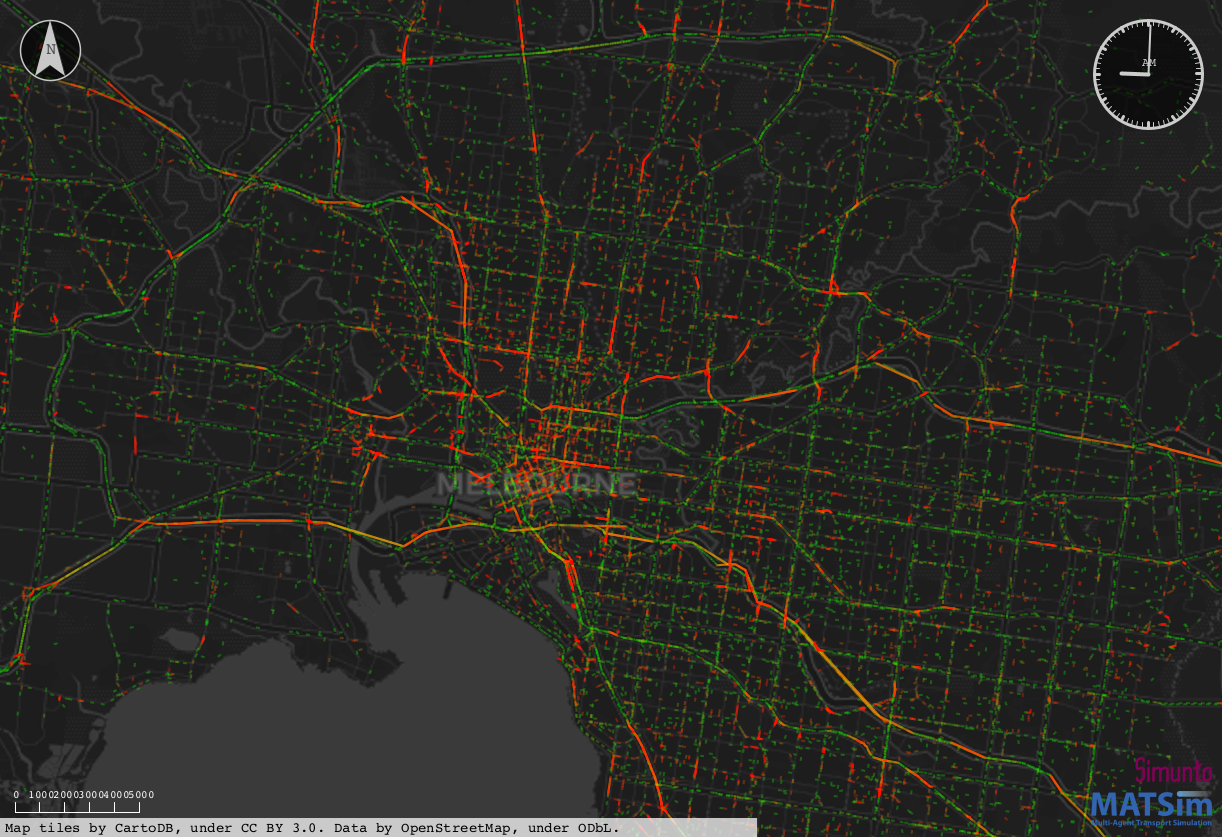} 
        	\caption{Morning peak at 9 AM}
        	\label{fig:amPeakVia}
        \end{subfigure}
        \begin{subfigure}[b]{0.47\textwidth}
        	\includegraphics[width=\textwidth]{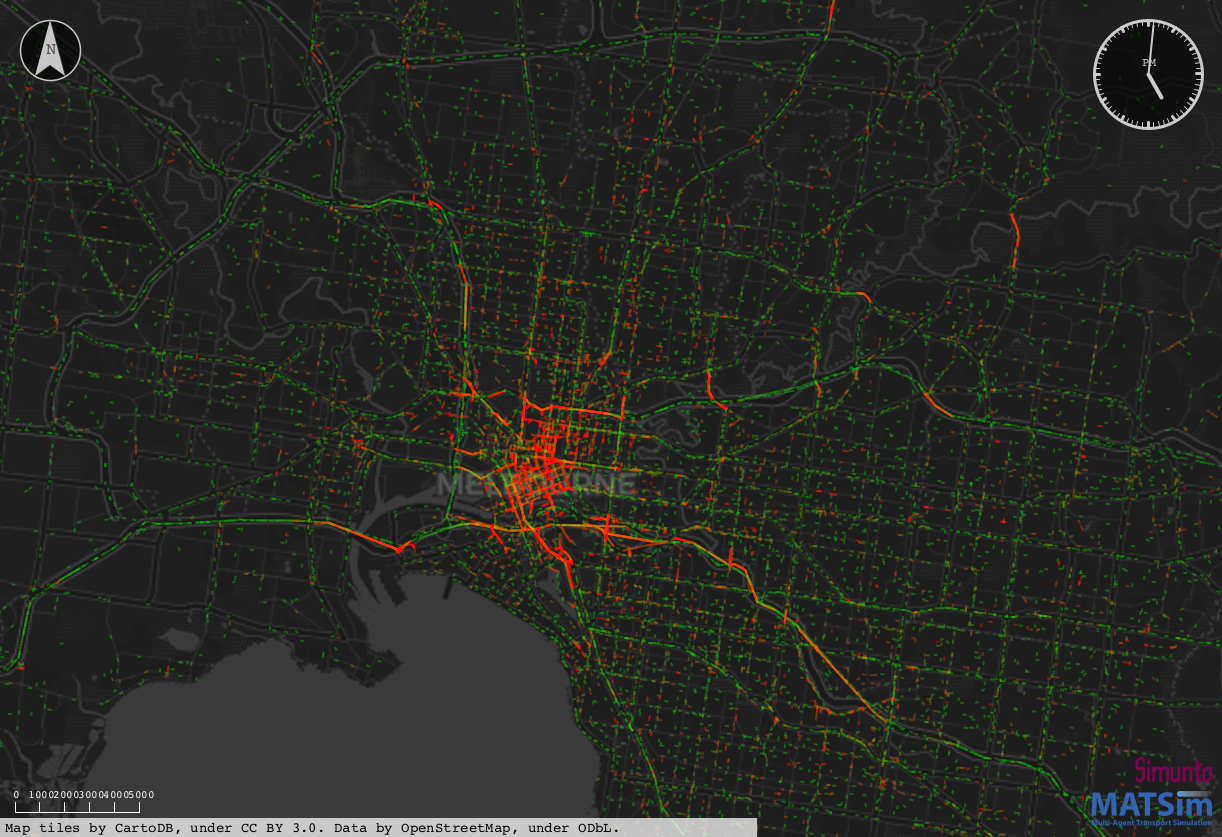} 
        	\caption{Evening peak at 5 PM}
        	\label{fig:pmPeakVia}
        \end{subfigure}
        \caption{Snapshots of the simulated car traffic at (a) morning peak (9:00 AM) and (b) evening peak (5:00 PM) for inner Melbourne. Colours represent the relative speed with red = full stop, yellow = travelling speed equal to half of the speed limit, green = travelling speed equal to the speed limit.}
        \label{fig:peaktrafficvia}
        \end{figure}

\begin{figure}[h]
        \centering
        \begin{subfigure}[b]{0.47\textwidth}
        	\includegraphics[width=\textwidth]{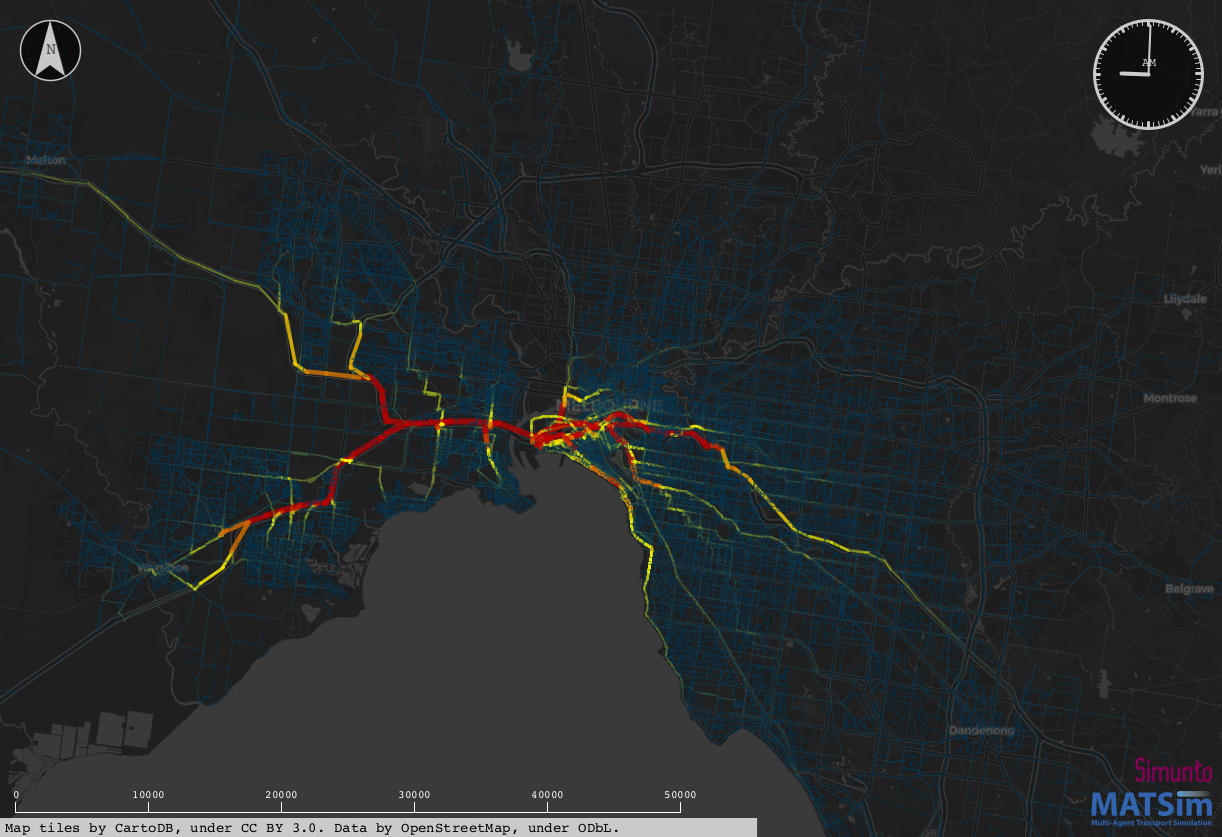} 
        	\caption{}
        	\label{fig:wgbFeeders}
        \end{subfigure}
        \begin{subfigure}[b]{0.47\textwidth}
        	\includegraphics[width=\textwidth]{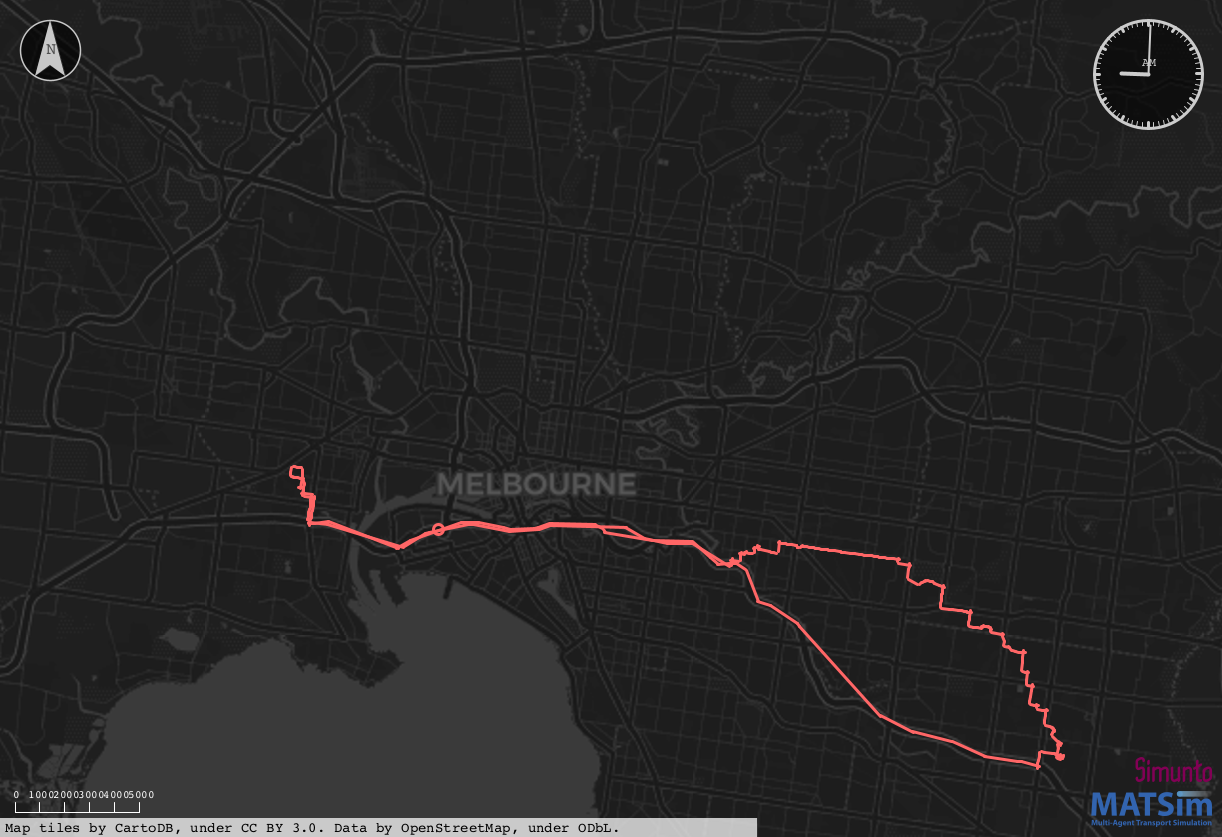} 
        	\caption{}
        	\label{fig:wgbDriver}
        \end{subfigure}
        \caption{West Gate Bridge 9:00 AM snapshot of (a) where the agents using it are coming from and travelling to and (b) simulated travel route of an example agent using the bridge.}
        \label{fig:westGatesBridge}
        \end{figure}

    In addition to mode choice, the car traffic volumes as well as the \gls{pt} passenger flow at the \gls{lga} level from the simulation model output also resemble the volumes observed in the real world, Figures~\ref{fig:carTotalVols} and \ref{fig:ptflow}, respectively.
    Figure~\ref{fig:timeDistDiff} shows that in addition to the road level and aggregated level, the model results also reflect the expected behaviour in terms of travel time and distance at the trip level.
    The realistic road traffic behaviour of the model makes it suitable for examining various traffic management interventions such as modifying speed limits or blocking certain roads to guide the traffic flow.
    For example, a snapshot of car traffic on roads within 10km radius of the Melbourne CBD for 9 AM and 5 PM is illustrated in Figure~\ref{fig:peaktrafficvia}, depicting heavy congestion on major roads connecting the Melbourne CBD to rest of the metropolitan area.
    Using agent-based models, it is possible to go beyond high-level snapshots and examine the road usage at the individual level. 
    For instance, one of the road segments with heavy congestion both in AM and PM peak hours is the West Gate Bridge, Victoria's most heavily used bridge, which is responsible for connecting the Melbourne CBD to the western suburbs.
    Figure~\ref{fig:wgbFeeders} illustrates where vehicles using this road segment at 9AM are coming from and heading to, confirming the critical role the bridge is playing in connecting the western suburbs to rest of the centre and to east.
    Furthermore, the travel route of an example agent using this bridge at 9 AM is also highlighted in Figure~\ref{fig:wgbDriver}.

    All categories of roads accessible to the public were included in the road network of the model, including minor bike paths to local streets and to major arterial roads and highways. 
    Therefore, in addition to common measures such as zone-to--to-zone movements or traffic on major highways and corridors, our model can be used for exploring local road usage for accessing local destinations.
    Further calibration for local road usage is needed to get reliable local road usage for active modes of transport from the model.    
        
    Unavailability of proper data on walking and cyclists' behaviour at the city scale has been a major barrier in designing interventions for promoting active transport.
    Typically available city-scale data for walking and cycling are limited to selected counting points or to a specific group of participants in a study or users of a smartphone application.
    Our model development workflow has the potential to fill this gap as it includes all road categories accessible to pedestrians and cyclists, attaches road attributes such as slope and bikeway type to the road network---even though these attributes were not included in the simulation model---and creates a synthetic population with individual attributes such as age, gender, occupation, and household structure, that are all important factors for walking and cycling behaviour.
    However, the current version of the simulation model does not consider the impact of these road and individual attributes on the travel behaviour (i.e., mode choice and route choice) of pedestrians and cyclists, which resulted in the comparatively high inaccuracies observed in the cycling and walking road usage analysis of Figure~\ref{fig:bikeTotalWape} and Figure~\ref{fig:walkTotalWape} when compared to the driving analysis.  
    Therefore, more research is needed to make the simulation model suitable for analysing active transport road usage.

    A key advantage of the workflow presented here is that it can also be integrated with other models.  For example, the simulation output analysis tools provided in our workflow convert the simulation outputs into formats (e.g., hourly traffic counts joined to the network, individual travel diaries, minutes spent walking) that are straightforward to join to existing models and tools. Health impact assessment tools such as the Transport Health Assessment Tool for Melbourne (THAT-Melbourne) \citet{gunn_helping_2021,zapata_transport_2021}, estimate non-communicable disease impacts \citep{zapata-diomedi_physical_2019,veerman_cost-effectiveness_2016} and some also include environmental impacts  such as air quality \citep{woodcock_health_2021} arising from travel behaviour change occurring between scenarios. A combined \gls{abm}-HIA model could be used to examine the health, economic, and environmental impacts of potential travel behaviour changes. 
    
    An important limitation of our model is that the mode choice parameters are estimated based on mandatory trips to work and education and mode choice is only permitted for workers.
    Further research is required to expand the mode choice model to include discretionary trips.
    Additionally, the model's travel mode choice function was only calibrated to match the percentages of work trips observed in Census 2016 and VISTA 2016-18.
    The calibration would have been strengthened, if the prediction accuracy of the model had been examined using a historical intervention; and this should be considered for future research if appropriate intervention data can be sourced.   
    Therefore, caution must be taken when interpreting the changes in mode share as a result of interventions and selecting the types of intervention to test using the model.
    
    Lastly, \gls{pt} trips were simulated based on deterministic timings from \gls{gtfs} and direct dedicated links connecting \gls{pt} stops (Figure~\ref{fig:ptNetwork}).
    Therefore, \gls{pt} vehicles had no interaction with other modes of transport while travelling and were strictly always on time.
    In reality, of course, tram and bus routes typically share the road with cars and are delayed due to traffic congestion or are a contributor to traffic congestion by occupying a significant amount of mixed traffic road space.
    In the current state of the model, neither of these two scenarios are captured, but again warrant future research. 
    
    In conclusion, this paper describes our open-source workflow for developing the \gls{atom} baseline scenario from building simulation model inputs to output post-processing.
    The model's mode choice coefficients were calibrated to capture the mode shares observed for trips to primary destinations in real-world for the four main travel models of driving, public transport, cycling, and walking.
    Furthermore, the comparison with real-world data showed that the model resembles peak hour car traffic volumes and \gls{pt} stations usage distributions observed in real-world as well as realistic travel times and distances.
    The model can be used to examine the potential impact of different scenarios on mode share, including active transportation and road traffic volume change.
   
\section*{Acknowledgements}
AJ is supported by an Australian Government Research Training Program Scholarship.
DS's time on this project is funded by Collaborative Research Project grants from CSIRO's Data61 (2018-19, 2020-21).
AB is supported by the NHMRC/UKRI JIBE project (\#APP1192788).
LG, MA, and SP are supported by the NHMRC funded Australian Prevention Partnership Centre (\#9100001); and BGC is supported by an RMIT VC Professorial Fellowship.

\bibliographystyle{apalike}
\bibliography{main.bib}
\end{document}